\documentclass[prb,reprint,aps,superscriptaddress,nofootinbib,longbibliography]{revtex4-2}
\usepackage{graphicx}
\usepackage{dcolumn}
\usepackage{bm}
\usepackage{braket}
\usepackage{amsmath}
\usepackage{amssymb}
\usepackage{feynmp}
\usepackage{graphics}
\usepackage{enumerate}
\usepackage{CJKutf8}
\usepackage[dvipsnames]{xcolor}
\usepackage{comment}
\usepackage[normalem]{ulem} 
\usepackage[hidelinks]{hyperref}

\renewcommand{\arraystretch}{1.5}
\everymath{\displaystyle}

\begin{document}

\title{A neural quantum state variational Monte Carlo study of Aharonov-Casher bands}

\author{Conor Smith\textsuperscript{*,\dag}}
\affiliation{Center for Computational Quantum Physics, Flatiron Institute, New York, New York 10010, USA}

\author{Nicol\'as Morales-Dur\'an\textsuperscript{*,\ddag}}
\affiliation{Center for Computational Quantum Physics, Flatiron Institute, New York, New York 10010, USA}

\author{Jingtian Shi}
\affiliation{Materials Science Division, Argonne National Laboratory, Lemont, Illinois 60439, USA}

\author{Andrew J. Millis}
\affiliation{Center for Computational Quantum Physics, Flatiron Institute, New York, New York 10010, USA}
\affiliation{Department of Physics, Columbia University, New York, New York 10027, USA}

\author{Miguel A. Morales}
\affiliation{Center for Computational Quantum Physics, Flatiron Institute, New York, New York 10010, USA}

\author{Shiwei Zhang}
\affiliation{Center for Computational Quantum Physics, Flatiron Institute, New York, New York 10010, USA}

\date{\today}

\begin{abstract}
We use neural-network-based variational Monte Carlo to investigate the competition between fractional Chern insulators and generalized Wigner crystals in the Aharonov-Casher Hamiltonian, a continuum model for moiré materials. We introduce a variational wave function inspired by flux attachment that captures fractionalized phases and, by comparing its energy with that of a Slater–Jastrow 
ansatz that captures crystalline phases, we determine the phase diagram at fractional fillings in regimes of strong band mixing and for system sizes beyond the reach of other methods. We find that discrete translation symmetry, encoded in a nonuniform quantum geometry, affects fractionalized states at filling fractions $\nu=1/3$ and $\nu=2/3$ differently, and that band mixing can stabilize incommensurate crystalline ground states. Our results further reveal the delicate competition between topological and non-topological crystalline phases.
\end{abstract}

\maketitle
\begingroup
\renewcommand{\thefootnote}{\fnsymbol{footnote}}
\footnotetext[1]{These authors contributed equally.}
\footnotetext[2]{\texttt{csmith@flatironinstitute.org}}
\footnotetext[3]{\texttt{nmoralesduran@flatironinstitute.org}}
\endgroup

\section{Introduction}

The interplay between wave function topology and electronic correlations gives rise to a rich variety of phases in moiré materials, including fractional Chern insulators (FCIs) \cite{cai2023signatures, zeng2023thermodynamic, park2023observation, xu2023observation,lu2024fractional}, topological crystals \cite{sun2026twistangle,Lu2025Extended,Aronson2025Displacement,xu2025superconductivity}, and superconductors \cite{xu2025superconductivity,Han2025Signatures}. Understanding the microscopic mechanisms driving the phase competition is an important scientific question, which is crucial for interpreting and guiding experimental efforts. However, the confluence of non-trivial quantum geometry, band dispersion and strong interactions poses a challenge for theoretical approaches. Controlled analytic results are only available in certain idealized limits \cite{wang2021exact,ledwith2020fractional,ledwith2023vortexability}, and existing numerical methods each face limitations. Hartree-Fock (HF) misses correlations beyond mean field, exact diagonalization (ED) is restricted to small systems and relies on projection onto a few bands, and 
similarly the density matrix renormalization group (DMRG) is limited in the number of bands and in its access to bulk two-dimensional systems.

In recent years, neural quantum state variational Monte Carlo (NQSVMC) \cite{Carleo2017Solving,Pfau2020Abinitio} has emerged as a powerful approach for studying the ground state of different correlated models, including the electron gas \cite{Pescia2024Message,Smith2024WC}, lattice models \cite{Viteritti2023Transformer,lange2026simulating,roth2025superconductivity,viteritti2026approaching,viteritti2026variational,roth2026largescale}, Landau levels \cite{Qian2025Describing,Teng2025Solving,Gattu2026Dressing,zhu2026crystallization,graham2026nonabelian,Perez2026Extracting,Abouelkomsan2026First,Abouelkomsan2026Topological}, and continuum models
for semiconductor moiré materials \cite{li2025deep,luo2025solving,smith2026neural,Geier2025Self,li2024emergent} and rhombohedral graphene \cite{valenti2025quantum,abouelkomsan2026rhombohedral}. For continuum Hamiltonians, real space NQS avoids explicit band projections and enables access to larger system sizes than those accessible to ED. Nevertheless, accurately representing topological order 
in the bulk limit remains a challenge for existing NQS approaches \cite{li2025deep,luo2025solving,Perez2026Extracting,Abouelkomsan2026First,Abouelkomsan2026Topological}.

In this work, we use NQSVMC to determine the phase diagram of the Aharonov-Casher (AC) Hamiltonian \cite{aharonov1979ground}, a model that contains the essential physics of semiconductor moiré materials such as tMoTe$_2$ \cite{Duran2024Magic,Shi2024Adiabatic}. The AC Hamiltonian describes a Schr\"odinger particle in a nonuniform magnetic field and gives rise to a flat low-energy band that retains the holomorphic structure of the lowest Landau level (LLL) even in the absence of continuous magnetic translation symmetry --an ideal band \cite{wang2021exact,ledwith2020fractional,ledwith2023vortexability}. By changing the spatial modulation of the magnetic field we can tune the band's quantum geometry, and by tuning the strength of electron-electron interactions we can control the amount of band mixing. This tunability makes the AC model a platform for studying the stability of fractionalized states and their competition with other phases.

We use two physics-informed neural-network variational wave functions: an ansatz inspired by flux attachment and
one in the Slater–Jastrow form. 
The flux-attachment ansatz accurately captures fractionalized phases as system size increases, addressing a key limitation encountered in previous studies, while the Slater–Jastrow ansatz accurately represents competing phases with crystalline order. We validate both wave functions by benchmarking against multiband ED on small systems. We then obtain the phase diagram at fractional fillings of the lowest AC band, in the presence of Coulomb interactions and for systems of up to 36 electrons, accessing regimes of strong band mixing at system sizes beyond the reach of existing methods. More generally, our variational framework provides an expressive and scalable approach that incorporates correlations and multiband physics in continuum models. 

We identify FCIs \cite{Wen_FCI,Neupert_FCI,DasSarma_Sun_FCI,Bernevig_Regnault_FCI}, as well as electron and hole generalized Wigner crystals (GWCs) \cite{Yoshioka_Fukuyama,MacDonald1985Broken,maki1983static} with Chern numbers $C=0$ and $C=1$, respectively. As the band's quantum geometry becomes more nonuniform, the FCI energy remains nearly constant at $\nu=1/3$, while decreasing rapidly at $\nu=2/3$. Sufficiently strong band mixing or Berry curvature concentration favors crystallization. Within the crystalline regime, electron and hole GWCs compete closely, and the ground state can be incommensurate with the periodicity of the magnetic field.

The remainder of this paper is organized as follows: In Sec.~\ref{sec:Hamiltonian} we discuss the AC Hamiltonian. In Sec.~\ref{sec:NQSVMC} we provide details on the variational wave functions employed. In Sec.~\ref{sec:Phase_Diagram} we present the results for the phase diagram at filling fractions $\nu=1/3$ and $\nu=2/3$. Finally, in Sec.~\ref{sec:Discussion} we 
conclude with a summary of our findings and discussions of some future directions.
 
\section{Interacting Hamiltonian}
\label{sec:Hamiltonian}
We solve for the ground state of the Hamiltonian 
\begin{align}
    H=H_{\text{AC}}+\frac{1}{2}\sum_{i\neq j} \frac{e^2}{\varepsilon|{\bm r}_i-{\bm r}_j|},
    \label{eq:Interacting_Hamiltonian}
\end{align}
where $i,j$ are particle indices, $\varepsilon$ is the dielectric constant and the single-particle piece is the Aharonov-Casher Hamiltonian \cite{aharonov1979ground}
\begin{align}
    H_{\text{AC}}=\frac{1}{2m^*}\left[ \bm \Pi^2+B({\bm r})\right],
    \label{eq:AC_Hamiltonian}
\end{align}
which describes a single spin sector of a non-relativistic particle of effective mass $m^*$, subject to a nonhomogeneous magnetic field $B({\bm r})$. The kinetic-momentum operator in Eq.~\eqref{eq:AC_Hamiltonian} is defined as $\bm \Pi={\bm p}+{\bm A}(\bm r)$, where $\nabla \times {\bm A}({\bm r})=B({\bm r})$. To make a connection with moiré materials, we choose the magnetic field to have triangular periodicity with lattice constant $a$, and split it as
\begin{align}
    B({\bm r})=B_0+\delta B({\bm r}).
    \label{eq:MagneticFieldSplit}
\end{align}
The average field $B_0$ is fixed so that each unit cell encloses one magnetic flux quantum, which determines the magnetic length  $\ell^2=\hbar/(eB_0)$. For concreteness, and in order to facilitate comparison with existing results, we take $\delta B({\bm r})$ to consist only of its first-shell harmonics:
\begin{align}
    \delta B({\bm r})=\sum_{{\bm G}}B_1\, e^{i\, {\bm G }\cdot{\bm r}},
\end{align}
where ${\bm G}$ are the first-shell reciprocal lattice vectors of the triangular lattice. Eq.~\eqref{eq:AC_Hamiltonian} gives rise to a flat lowest-energy $C=1$ band for any value of the harmonic $B_1$, which has units of $\ell^{-2}$ and is taken to be real.

\begin{figure*}[htbp]
    \centering
    \includegraphics[width=\textwidth]{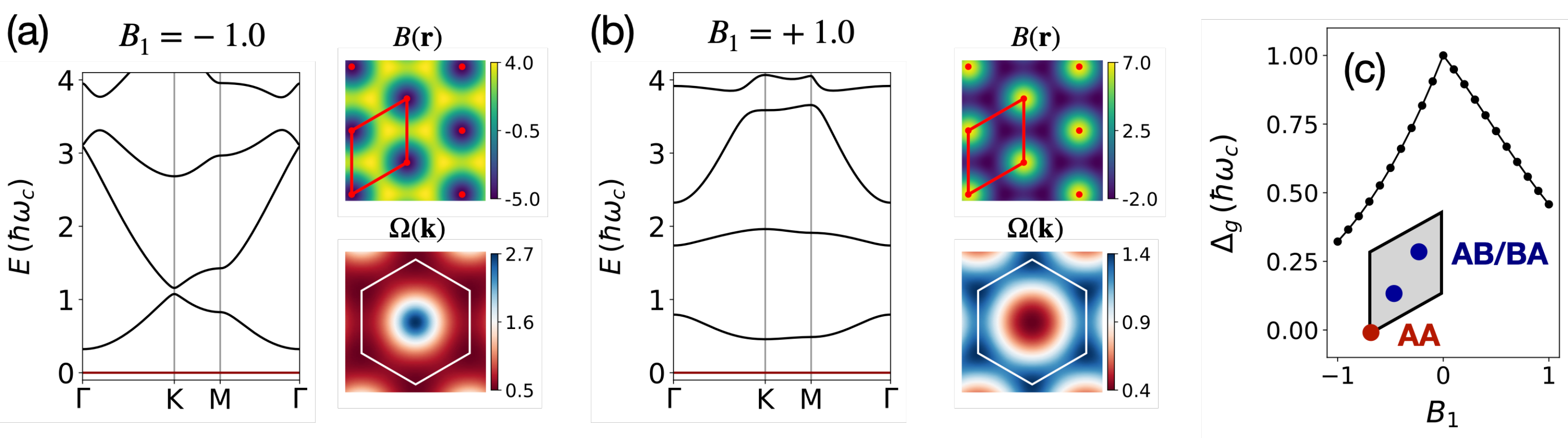}
    \caption{Band structure of the Aharonov-Casher Hamiltonian, Eq.~\eqref{eq:AC_Hamiltonian}, for parameters (a) $B_1=-1.0$ and (b) $B_1=+1.0$. The lowest band (indicated in red) is flat and its wave functions are connected to those of the lowest Landau level. The magnetic field $B({\bm r})$ giving rise to each band structure and the Berry curvature $\Omega({\bm k})$ of the flat band are also shown. The unit cell and the Brillouin zone are indicated. (c) Band gap between the first two AC bands as a function of $B_1$. The inset indicates the high-symmetry points within the unit cell of the magnetic field, which we label AA and AB/BA. All energies are measured in units of the cyclotron gap in the uniform field limit, $B_1=0$. }
    \label{fig:BandStructure}
\end{figure*}

For $ B_1=0$, the magnetic field is uniform and Eq.~\eqref{eq:AC_Hamiltonian} reduces to the Landau levels of a non-relativistic particle of mass $m^*$. When $B_1\neq0$, the lowest-energy band from Eq.~\eqref{eq:AC_Hamiltonian} is still flat and retains the holomorphic structure of the LLL \cite{aharonov1979ground,Duran2024Magic,Shi2024Adiabatic,sommer2026ideal}, but the remote bands become dispersive. Representative band structures resulting from Eq.~\eqref{eq:AC_Hamiltonian} are shown in Fig.~\ref{fig:BandStructure}(a),(b) for two values of $B_1$. The periodic magnetic fields from which the band structures emerge and the resulting Berry curvature of the flat band are also shown in Fig.~\ref{fig:BandStructure}(a),(b). The evolution of the band gap $\Delta_g$ between the two lowest AC bands as a function of $B_1$ is shown in Fig.~\ref{fig:BandStructure}(c).

Here we consider simple magnetic field distributions given by Eq.~\eqref{eq:MagneticFieldSplit}, whose extrema are at the high-symmetry points within the unit cell, which we label AA and AB/BA, as shown in the inset of Fig.~\ref{fig:BandStructure}(c). 
The parameter $B_1$ controls the degree of nonuniformity of the flat band’s Berry curvature $\Omega(\bm k)$ and the location of its extrema in the Brillouin zone. Therefore, Eq.~\eqref{eq:Interacting_Hamiltonian} provides a simple model to study many-body physics in flat Chern bands with nonuniform Berry curvature, which are common in moiré materials. In fact, the lowest-energy band of Eq.~\eqref{eq:AC_Hamiltonian} can be related to the topmost valence band of the continuum Hamiltonian for twisted transition metal dichalcogenide homobilayers \cite{Wu2019Topological} through the adiabatic approximation. See Refs. \cite{Duran2024Magic,Shi2024Adiabatic} for further details.

Previous work has shown that the AC model at fillings $\nu<1$ can support FCIs \cite{shi2025effects,iyer2026dispersion,moralesduran2026bandmixing}, generalized Wigner crystals ~\cite{Duran2026nonuniform}, and even superconductivity (for a Haldane pseudopotential model interaction)\cite{Guerci_AC_SC}. However, these studies rely on projection onto one or two bands and are constrained by system size. From Fig.~\ref{fig:BandStructure}(c) we see that for a projection onto the lowest band to be justified, the interaction strength should be smaller than the gap, which limits the range of validity of band-projected studies. Here we overcome those limitations by studying the AC Hamiltonian, with Coulomb interactions, using neural quantum state variational Monte Carlo. 

We obtain the phase diagram at fractional fillings of the flat band, as a function of quantum geometry nonuniformity (controlled by $B_1$) and band mixing (controlled by the strength of interactions). In our calculations, the interaction strength is controlled by the dielectric constant $\varepsilon$ and quantified by the Wigner-Seitz parameter:
\begin{align}
    r_s=\sqrt{\frac{2}{\nu}}\,\left(\frac{e^2/\varepsilon\ell}{\hbar\omega_c}\right) =\sqrt{\frac{2}{\nu}}\,\kappa,
\end{align}
where $\nu$ is the filling fraction and $\hbar \omega_c=\hbar e B_0/m^*$ is the cyclotron gap associated with the average field $B_0$. The second equality relates $r_s$ to the conventional Landau level mixing parameter $\kappa$. At fixed $B_1$ , the limit $r_s\to0$ corresponds to projection of the Coulomb interaction onto the lowest AC band.

\section{Neural Quantum State Variational Monte Carlo}
\label{sec:NQSVMC}
We consider a finite system containing $N$ unit cells of the magnetic field on a torus of dimensions $L_x\times L_y$, with periodic boundary conditions. The system is threaded by $N_\phi=N$ flux quanta \cite{Duran2024Magic}. We solve for the ground state of $N_e$ electrons, such that the filling fraction is $\nu=N_e/N$.

\subsection{Generalized orbitals}
Both variational wave functions used in this work contain a Slater determinant of generalized orbitals of the form
\begin{equation}
\phi_a(\bm r_i;\{\bm r\})=\eta_a(\bm r_i;\{\bm r\})\sum_{k_m}\sum_{n=0}^{n_{\text{max}}}
   c^{nk_m}_{a}\,\psi^{(N_\phi')}_{nk_m}(\bm r_i),
    \label{eq:landau_orbitals}
\end{equation}
where $\psi^{(N_\phi')}_{nk_m}$ is an $n$-th Landau level orbital associated to a uniform magnetic field threading the torus with $N_\phi'$ flux quanta. The momentum $k_m=2\pi m/L_x$, with $m=0,\cdots, N_{\phi}^{\prime}-1$ labels the guiding center, while $n_{\text{max}}$ is the cutoff that determines the basis size. The multiplicative factor $\eta_a(\bm r_i;\{\bm r\})$ is the output of an all-body neural network and $c^{nk_m}_{a}$ are variational parameters. The two ansätze considered in this work differ in the flux $N_\phi'$ carried by the orbitals, {\it i.e.}, in how the total flux $N_\phi$ is distributed among the factors of the wave function. We now introduce the two neural ans\"atze.

\subsection{Slater–Jastrow ansatz}
In the ``Landau" Slater–Jastrow (LSJ) ansatz the orbitals carry all of the flux, namely $N_\phi'=N_\phi$, and
\begin{align}
    \Psi_{\text{LSJ}}(\{\bm r\})=J(\{\bm r\})\,\det_{i,a}\big[\phi_a(\bm r_i;\{\bm r\})\big],
    \label{eq:SJ_Ansatz}
\end{align}
where $J(\{\bm r\})$ is a neural-network Jastrow factor. Equation~\eqref{eq:SJ_Ansatz} can accurately describe crystalline and metallic states, and it is also capable of representing fractional quantum Hall correlations \cite{Abouelkomsan2026First}. However, learning these correlations becomes increasingly difficult as the system size grows \cite{Perez2026Extracting,Abouelkomsan2026First}. 

\subsection{Composite-fermion ansatz}
In order to express the nontrivial correlations of topologically ordered states we consider a generalization of the Jain wave function \cite{Jain1989Composite,Pu2017Composite,Zhao2018Crystallization},
\begin{equation}
\begin{split}
\Psi_{\mathrm{CF}}(\{\bm r\}) = {}& J(\{\bm r\})\,
\Phi_{2p}(\{ \bm r \})\det_{i,a}\!\big[\phi^{(\star)}_{a}(\bm r_i;\{\bm r\})\big] ,
\end{split}
\label{eq:CF_Ansatz}
\end{equation}
where  the flux-attachment factor is
\begin{equation}
\Phi_{2p}(\{\bm r\})=e^{-\sum_i y_i^2/2\ell_J^2}
\prod_{i<j}\vartheta_1(z_i-z_j|\tau)^{2p}\;\Theta_{\mathrm{CM}}(Z|\tau).
\label{eq:flux_attachment}
\end{equation}
Here $\vartheta_1$ is the Jacobi theta function, and $\Theta_{\mathrm{CM}}$ is a center-of-mass factor that selects a topological sector. The complex coordinate is defined as $z_i=(x_i+iy_i)/L_x$, where we have assumed that our simulation cell is rectangular and $L_x$ is the length of a single side, $Z=\sum\nolimits_i z_i$ is the center-of-mass coordinate and $\tau=iL_y/L_x$ is the modular parameter. The flux-attachment factor attaches $2p$ flux quanta to each electron, so the orbitals in the determinant see the reduced flux
\begin{equation}
N_\phi' = N_\phi^* = N_\phi - 2pN_e,
\label{eq:CF_flux}
\end{equation}
with a corresponding effective magnetic length given by $\ell_J=\sqrt{L_xL_y/2\pi N_\phi'}$.

At $\nu=1/3$, taking $p=1$ gives $N_\phi^*=N_e$, so the composite fermions fill one effective Landau level ($\nu^*=1$). At $\nu=2/3$, $p=1$ gives $N_\phi^*=-N_e/2$: the effective field is reversed and the composite fermions fill two effective Landau levels ($\nu^*=-2$). We account for the reversed field by complex conjugating the orbitals, denoted $\phi^{\star}_a$. For $\nu=1/m$, with $m$ an odd integer, this ansatz recovers the Laughlin states \cite{Laughlin1983Anomalous}. Although this wave function is designed to recover Jain states in a uniform magnetic field, it can also develop crystalline correlations due to the generalized orbitals. Details about the neural factors $\eta_a({\bm r}_i; \{{\bm r}\})$ and $J(\{\bm r\})$ are provided in Appendix~\ref{Appendix:Ansatz}.

Using both ansätze, we can describe the competing phases accurately for system sizes significantly larger than previously accessible with NQS in continuum magnetic systems on the torus~\cite{Abouelkomsan2026First,Perez2026Extracting}. To confirm the accuracy of our method, we benchmark our results against multiband ED on small systems.

\section{Phase diagram of the AC model}
\label{sec:Phase_Diagram}

In this work we focus on two filling fractions: $\nu=1/3$, which hosts a Laughlin-like state and for which previous results provide a benchmark; and $\nu=2/3$, where the most robust FCI states emerge in tMoTe$_2$. At both fillings, charge-ordered phases that spontaneously break the translation symmetry of the Hamiltonian compete with the FCIs. For $B_1\neq0$, this symmetry is discrete, and we refer to the charge-ordered phases as generalized Wigner crystals. This distinguishes them from conventional Wigner crystals in the uniform field limit, which break continuous magnetic translation symmetry.

\subsection{Results at $\nu=1/3$ filling}
We start by discussing the expected ground states for different limiting cases of the Hamiltonian given by Eq.~\eqref{eq:Interacting_Hamiltonian}. For the Landau level Hamiltonian ($B_1=0$) we expect the Laughlin state to be the ground state for small $r_s$, and an electron Wigner crystal with lattice constant $a_c=\sqrt{3}\, a$ to be the ground state for large $r_s$ \cite{Platzman1993QuantumFreezing,Yi1998Laughlin}. When interactions are projected to the lowest AC band ($r_s\to 0$), large magnetic field modulations favor an electron generalized Wigner crystal with lattice constant $a_c=\sqrt{3}\, a$, which is commensurate with the periodicity of the magnetic field \cite{Duran2026nonuniform}. Therefore, for both strong band mixing and large quantum geometric nonuniformity we expect the Laughlin FCI to be unstable against a crystalline ground state.

\begin{figure}[htbp]
    \centering
    \includegraphics[width=0.48\textwidth]{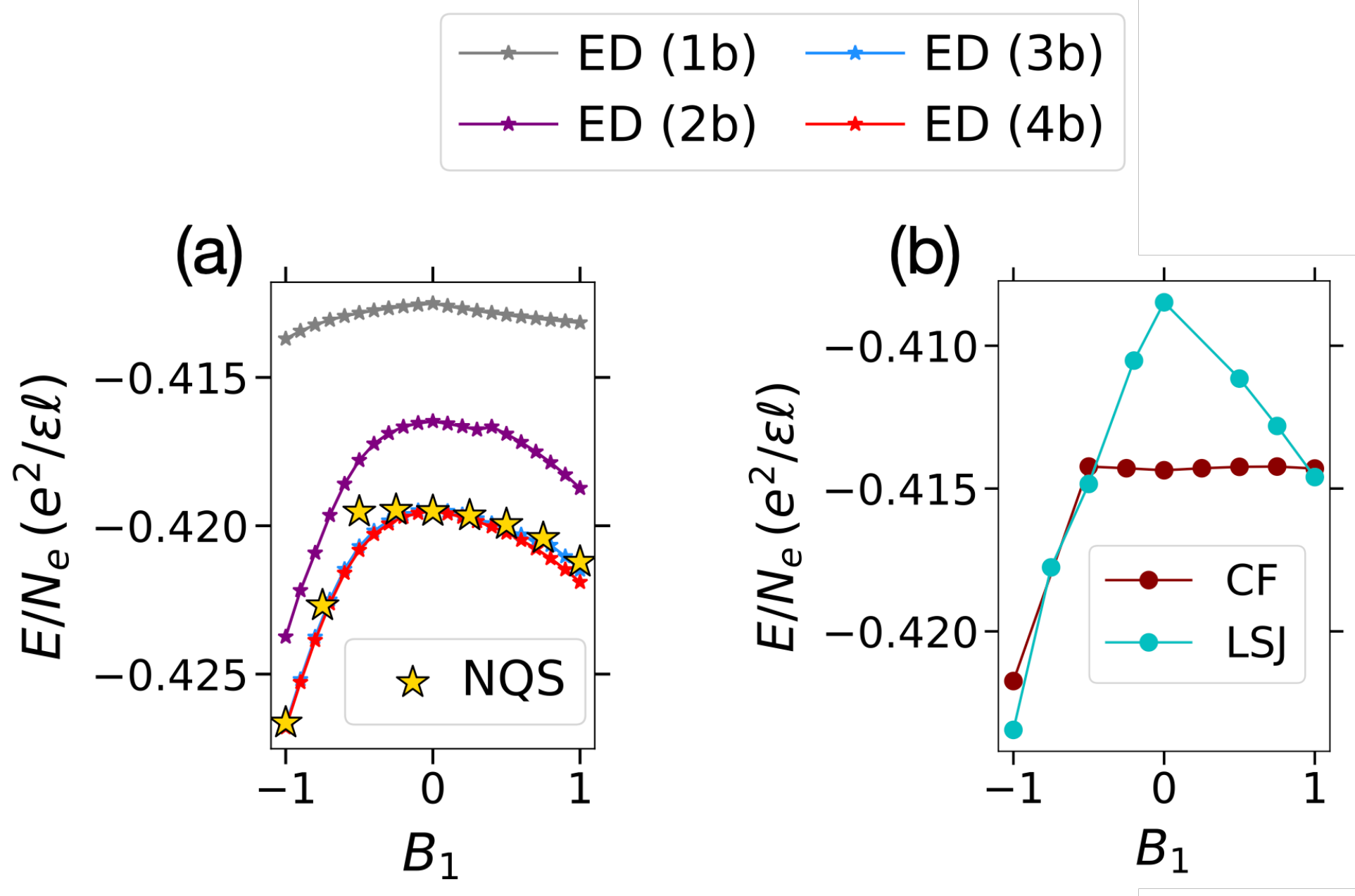}
    \caption{(a) Comparison between multiband ED and NQSVMC energies per particle as a function of $B_1$, for a system with $N_e=4$ electrons on $N=12$ unit cells ($\nu=1/3$). (b) Ground state energies per particle as a function of $B_1$, obtained from our two variational wave functions on a system with $N_e=36$ electrons on $N=108$ unit cells. For both plots we have used $r_s=9.8$.}
    \label{fig:nu_13_Energies}
\end{figure}

In Fig.~\ref{fig:nu_13_Energies}(a) we compare ED and NQSVMC ground state energies per particle as a function of $B_1$, for a system with $N_e=4$ electrons on $N=12$ unit cells. We include up to four bands in the Hilbert space of ED calculations, convergence is reached at the three-band level. NQSVMC reproduces the converged multiband ED energies and can be extended to larger system sizes.

In Fig.~\ref{fig:nu_13_Energies}(b) we compare the variational energies of the CF and LSJ ans\"atze as a function of $B_1$, for a system with $N_e=36$ electrons on $N=108$ unit cells. The CF wave function has lower energy for a range of $B_1$ around the Landau level limit ($B_1=0$). As $|B_1|$ increases, the energy of $\Psi_{\mathrm{CF}}$ remains almost constant, while the energy of $\Psi_{\mathrm{LSJ}}$ decreases roughly linearly and eventually falls below that of $\Psi_{\mathrm{CF}}$. These energy crossings signal phase transitions.

\begin{figure}[htbp]
    \centering
    \includegraphics[width=0.42\textwidth]{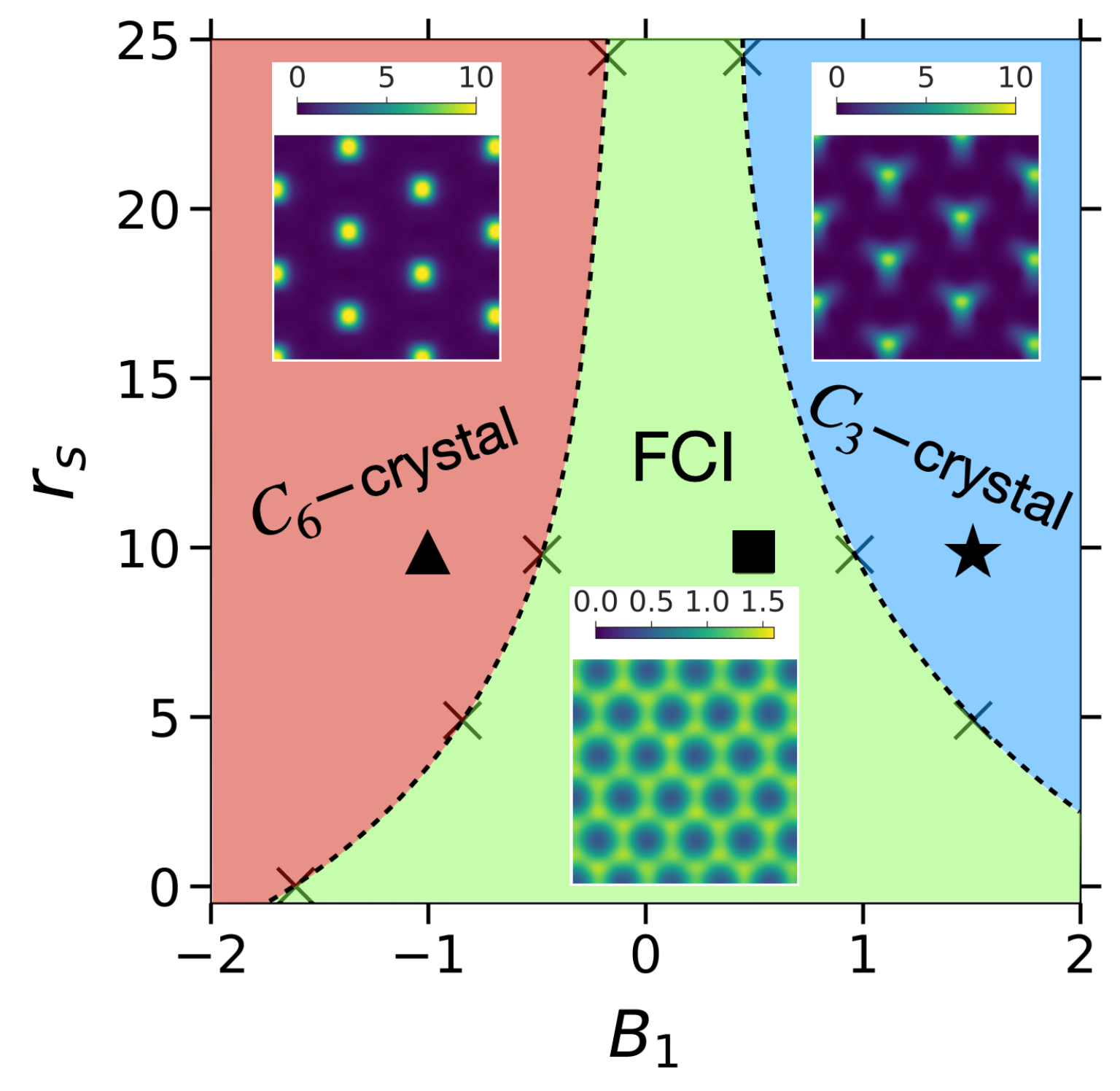}
    \caption{Schematic phase diagram of the AC Hamiltonian at $\nu=1/3$ as a function of $B_1$ and $r_s$. We identify a topologically ordered state with the Laughlin correlations and two electron generalized Wigner crystals. Representative charge densities for each phase are shown in the insets, the symbols indicate the parameters used to compute them. The FCI inherits the periodicity of the magnetic field, while the crystals break it to form a commensurate $\sqrt{3}\times\sqrt{3}$ order. The two crystals differ by the symmetry and location of their orbitals: The $C_6$-crystal has orbitals centered at AA and the $C_3$-crystal has orbitals centered at AB or BA. All the calculations were done for a system with $N_e=36$ electrons on $N=108$ unit cells. At $B_1=0$ the Laughlin state transitions into a crystal at $r_s\approx56.3$.}
    \label{fig:nu_13_PhaseDiagram}
\end{figure}

Figure~\ref{fig:nu_13_PhaseDiagram} shows the phase diagram of the AC Hamiltonian at $\nu=1/3$, as a function of $B_1$ and interaction strength $r_s$. We identify an FCI and two crystalline phases, which are consistent with those found in a previous study of the AC model at $\nu=1/3$ \cite{Duran2026nonuniform}. We characterize each phase by computing its charge density, structure factor, pair-correlation function and Chern number. The insets in Fig.~\ref{fig:nu_13_PhaseDiagram} show the ground state charge density at a representative state within each phase. Results for the remaining observables are presented in Appendix~\ref{Appendix:AdditionalResults}.

The FCI retains the Laughlin correlations (see Appendix~\ref{Appendix:AdditionalResults}), but its density becomes modulated with the periodicity of the magnetic field. The crystalline phases spontaneously break the discrete translation symmetry of the Hamiltonian to form a commensurate $\sqrt{3}\times\sqrt{3}$ charge order, namely a GWC. Both crystals are topologically trivial (they have Chern number $C=0$, see Appendix~\ref{Appendix:Ansatz}), but differ in the rotational symmetry of their orbitals. This difference reflects the distinct localization centers favored by the nonuniform magnetic field: Orbitals centered at AA sites inherit their $C_6$-symmetry, while orbitals centered at AB or BA sites inherit their $C_3$-symmetry \cite{Duran2026nonuniform}. Following previous work, we refer to the GWCs we find at $\nu=1/3$ as the $C_6$-crystal and the $C_3$-crystal \cite{mishra2026charge}.

\begin{figure}[htbp]
    \centering
    \includegraphics[width=0.46\textwidth]{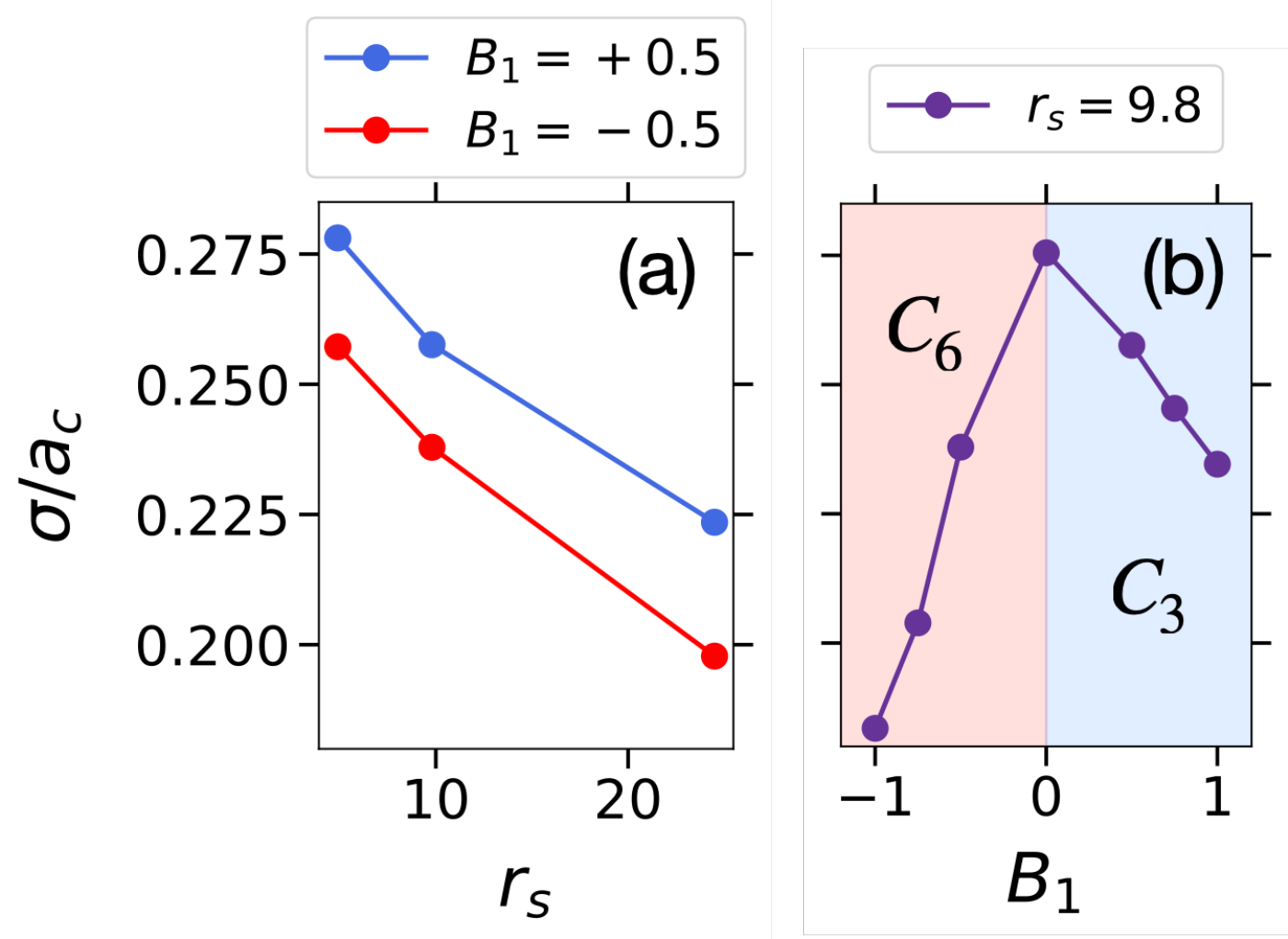}
    \caption{(a) Evolution of the normalized orbital spread $\sigma/a_c$ for the $C_6$-crystal ($B_1=-0.5$) and the $C_3$-crystal ($B_1=+0.5$) as a function of $r_s$. (b) Orbital spread of the electron GWC with the lowest-energy, as a function of $B_1$. The $C_6$-crystal has lower energy for $B_1<0$ and the $C_3$-crystal has lower energy for $B_1>0$.}
    \label{fig:nu_13_Lindemann}
\end{figure}

Crystallization is favored by nonuniform quantum geometry and band mixing, which can be understood in terms of orbital localization. To quantify this effect, we compute the orbital spread around the crystal sites ${\bm R}_j$,
\begin{equation}
    \sigma^2=\frac{1}{N_e}\sum_j\int d^2r\,({\bm r}-{\bm R_j})^2\,\rho({\bm r}),
\end{equation}
where $\rho({\bm r})$ is the ground state density, and the integration is over the crystal unit cell. Fig.~\ref{fig:nu_13_Lindemann} shows that the normalized spread $\sigma/a_c$, where $a_c$ is the crystal lattice constant, decreases with increasing $|B_1|$ and $r_s$, demonstrating enhanced localization. The energetic benefit from localization follows from a multipole expansion in the dilute limit: The leading finite-spread correction to the classical point-charge crystal energy is proportional to $\sum\nolimits_j\sigma^2/|{\bm R}_j|^3$ \cite{Duran2026nonuniform}. For the additional localization enabled by band mixing, the Coulomb energy gain outweighs the kinetic cost of the occupation of higher bands, lowering the total energy of the crystal.

Finally, we find that for the system with $N_e=36$ electrons the {\it freezing} of the Laughlin state at $B_1=0$ happens at $r_s\approx 56.3$, which corresponds to $\kappa\approx 23$, and is outside the range of the phase diagram in Fig.~\ref{fig:nu_13_PhaseDiagram}. A recent study with $N_e=12$ electrons has reported the transition value to be between $\kappa= 15$ and $\kappa= 20$ \cite{Abouelkomsan2026First}. The discrepancy between the results is likely due to finite size effects and the use of a slightly different variational ansatz.

\subsection{Results at $\nu=2/3$ filling}
In the limit $B_1=r_s=0$ the Hamiltonian in Eq.~\eqref{eq:Interacting_Hamiltonian} reduces to the Coulomb interaction projected to the LLL, whose particle-hole symmetry relates the ground states at fillings $\nu$ and $1-\nu$. Away from this limit, this equivalence is lost \cite{Lauchli2013Hierarchy,Abouelkomsan2023Quantum} and the phase diagram at $\nu=2/3$ can differ substantially from that at $\nu=1/3$. Crystalline states are still expected to dominate at $\nu=2/3$ for large $r_s$ and large $|B_1|$, but the phase competition at intermediate values of band mixing and quantum geometry nonuniformity has not been previously studied.

\begin{figure}[htbp]
    \centering
    \includegraphics[width=0.48\textwidth]{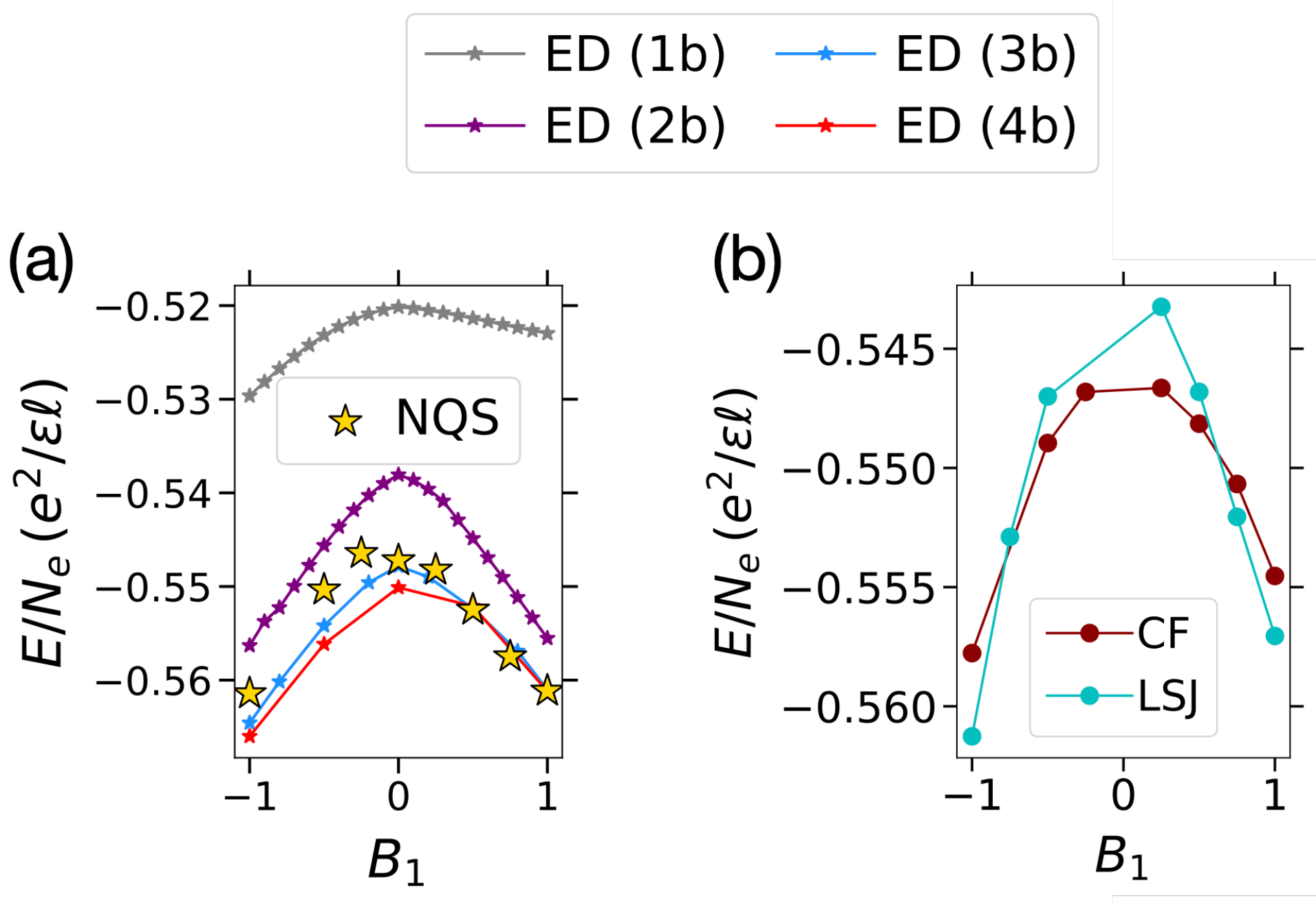}
    \caption{(a) Comparison between multiband ED and NQSVMC energies per particle as a function of $B_1$, for a system with $N_e=8$ electrons  on $N=12$ unit cells ($\nu=2/3$). (b) Ground state energies per particle as a function of $B_1$, obtained from our two variational wave functions on a system with $N_e=24$ electrons on $N=36$ unit cells. For both plots we have used $r_s=6.9$.}
    \label{fig:nu_23_Energies}
\end{figure}

In Fig.~\ref{fig:nu_23_Energies}(a) we compare ED and NQSVMC ground state energies as a function of $B_1$, for a system with $N_e=8$ electrons on $N=12$ unit cells. The ED energies are nearly converged at the four-band level. The agreement is good, even though the NQSVMC performance for small negative $B_1$ values is not as perfect as in the $\nu=1/3$ case, reflecting the more challenging nature of this system.
Fig.~\ref{fig:nu_23_Energies}(b) shows NQSVMC variational energies as a function of $B_1$ for a system with $N_e=24$ electrons on $N=36$ unit cells, which is beyond the reach of multiband ED. The CF ansatz has lower energy than the LSJ ansatz at $B_1=0$ but their energies cross for sufficiently strong magnetic field modulations, indicating phase transitions. 

\begin{figure}[htbp]
    \centering
    \includegraphics[width=0.48\textwidth]{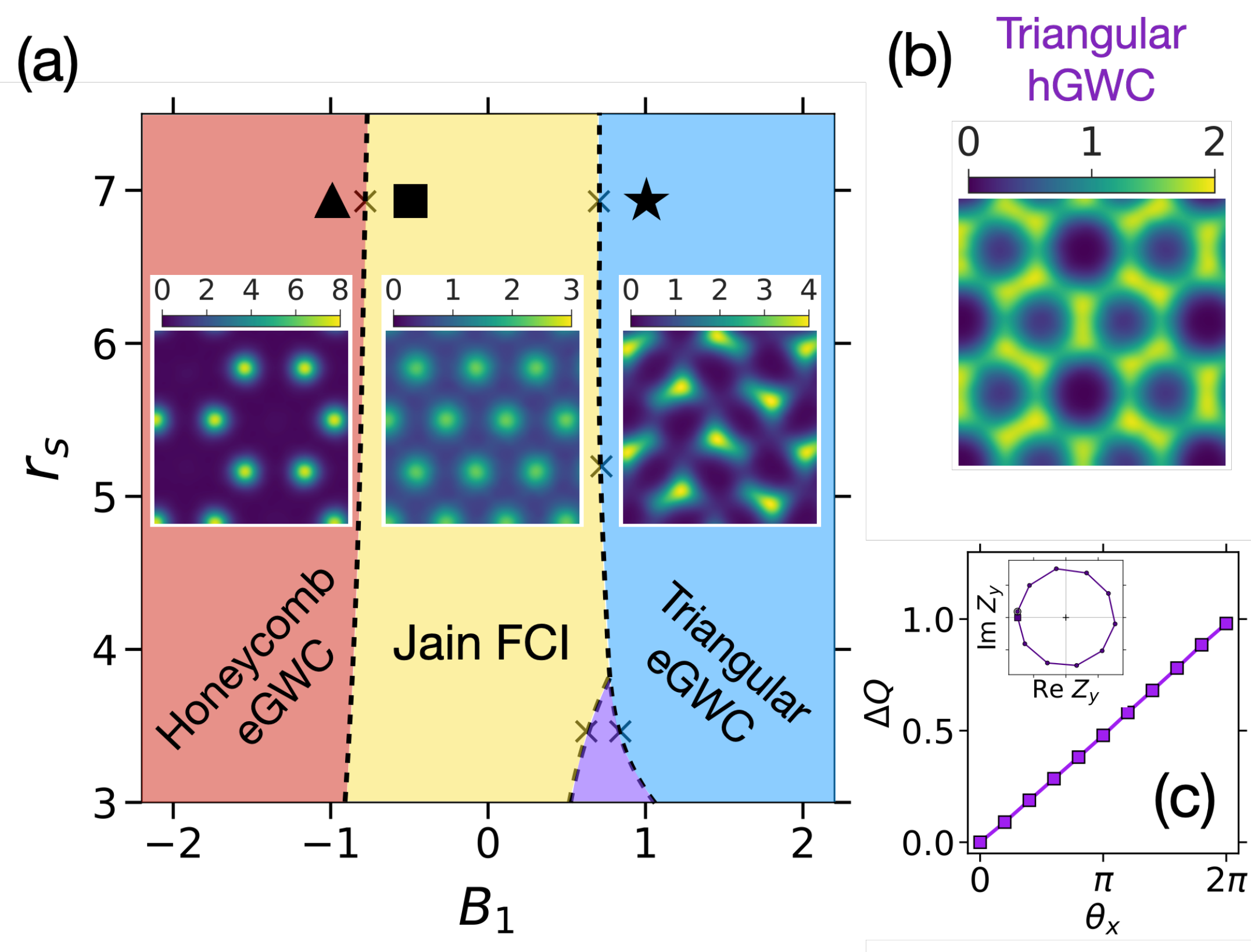}
    \caption{(a) Schematic phase diagram of the AC Hamiltonian at $\nu=2/3$ as a function of $B_1$ and $r_s$. We find a topologically ordered state with Jain correlations and three crystalline phases: a honeycomb electron GWC, a triangular (incommensurate) electron GWC, and a triangular hole GWC. Insets show representative charge densities for the two electron crystals and the Jain FCI, computed for the parameters indicated by the symbols. The triangular hGWC is shown separately in panels (b) and (c). (b) Charge density of the hole GWC at $B_1=+0.75$ and $r_s=3.46$. (c) Charge pumping from threading flux through the torus, confirming that the hole GWC has Chern number $C=+1$. The inset shows the phase winding. All the calculations were done for a system with $N_e=24$ electrons $N=36$ unit cells. At $B_1=0$ the Jain state transitions into a crystal at $r_s\approx44.5$.}
    \label{fig:nu_23_PhaseDiagram}
\end{figure}

In Fig.~\ref{fig:nu_23_PhaseDiagram}(a) we show the phase diagram of the AC Hamiltonian at $\nu=2/3$, as a function of $B_1$ and $r_s$. By computing charge densities, structure factors, pair-correlations functions, and Chern numbers, we identify four phases: an FCI with Jain-like correlations, and three crystalline states. In Fig.~\ref{fig:nu_23_PhaseDiagram} we show charge densities representative of each phase, and in Appendix~\ref{Appendix:AdditionalResults} we present results for the other observables.

The Jain-like FCI is the ground state around $B_1=0$, as expected from the connection of the AC model with Landau levels. Its density follows the periodicity of the magnetic field. We note that the energy of the Laughlin-like FCI at $\nu=1/3$ varies weakly with $B_1$, with leading corrections of order $B_1^2$ \cite{Duran2026nonuniform}, while the energy of the Jain-like FCI at $\nu=2/3$ varies more strongly with $B_1$. This indicates that these two fractionalized states respond differently to the breaking of continuous translational symmetry induced by the periodic magnetic field.

For large negative $B_1$, we find a commensurate honeycomb electron GWC with $C_6$-symmetric orbitals localized at the AA sites. For large positive $B_1$, the ground state is a triangular electron GWC with periodicity $a_c=\sqrt{3/2}\,a$ \cite{maki1983static}, which is adiabatically connected to the electron Wigner crystal in the uniform field limit, $B_1=0$. This periodicity is incommensurate with that of the magnetic field, which results in the density shown in Fig.~\ref{fig:nu_23_PhaseDiagram}(a). We find that the honeycomb electron GWC and the triangular electron Wigner crystal have Chern number $C=0$, which is consistent with their description in terms of localized electrons.

At intermediate positive $B_1$ and small $r_s$, we find a commensurate triangular hole GWC ground state. This phase is obtained by fully filling the lowest AC band with electrons ($\nu=1$) and introducing holes \cite{MacDonald1985Broken} that localize at the magnetic-field maxima. The localized holes do not contribute to transport, hence the state retains the Chern number of the filled band, $C=1$. In Fig.~\ref{fig:nu_23_PhaseDiagram}(b) we show the density of the hole GWC, which shows a triangular lattice of holes at the AA sites. Fig.~\ref{fig:nu_23_PhaseDiagram}(c) shows the charge pumped by flux threading, which confirms that the phase has Hall conductivity $\sigma_{xy}=e^2/h$. Hole Wigner crystals have been used to explain the re-entrant integer quantum Hall states observed in GaAs \cite{Li2010Observation,Liu2012Observation}, and are also known as {\it partial Hall crystals} \cite{Tesanovic1989Hall}.

The competition between the triangular electron GWC and triangular hole GWC at $B_1>0$ is interesting because the incommensurate electron crystal dominates most of this region. At $B_1=0$, the ground state is the Jain state, and the hole Wigner crystal has lower energy than the electron Wigner crystal \cite{MacDonald1985Broken}. Because the hole WC is commensurate with the periodic magnetic field, we expect a hole GWC, adiabatically connected to the hole Wigner crystal at $B_1=0$, to become the ground state once the field modulation destabilizes the Jain state. Early Hartree-Fock calculations on Landau levels showed that increasing band mixing lowers the energy of the electron WC below that of the hole WC \cite{MacDonald1984Influence}. Similarly, we find that at finite $B_1>0$ the hole GWC transitions into the electron GWC as $r_s$ increases. This transition is driven by electron localization.

\begin{figure}[htbp]
    \centering
    \includegraphics[width=0.45\textwidth]{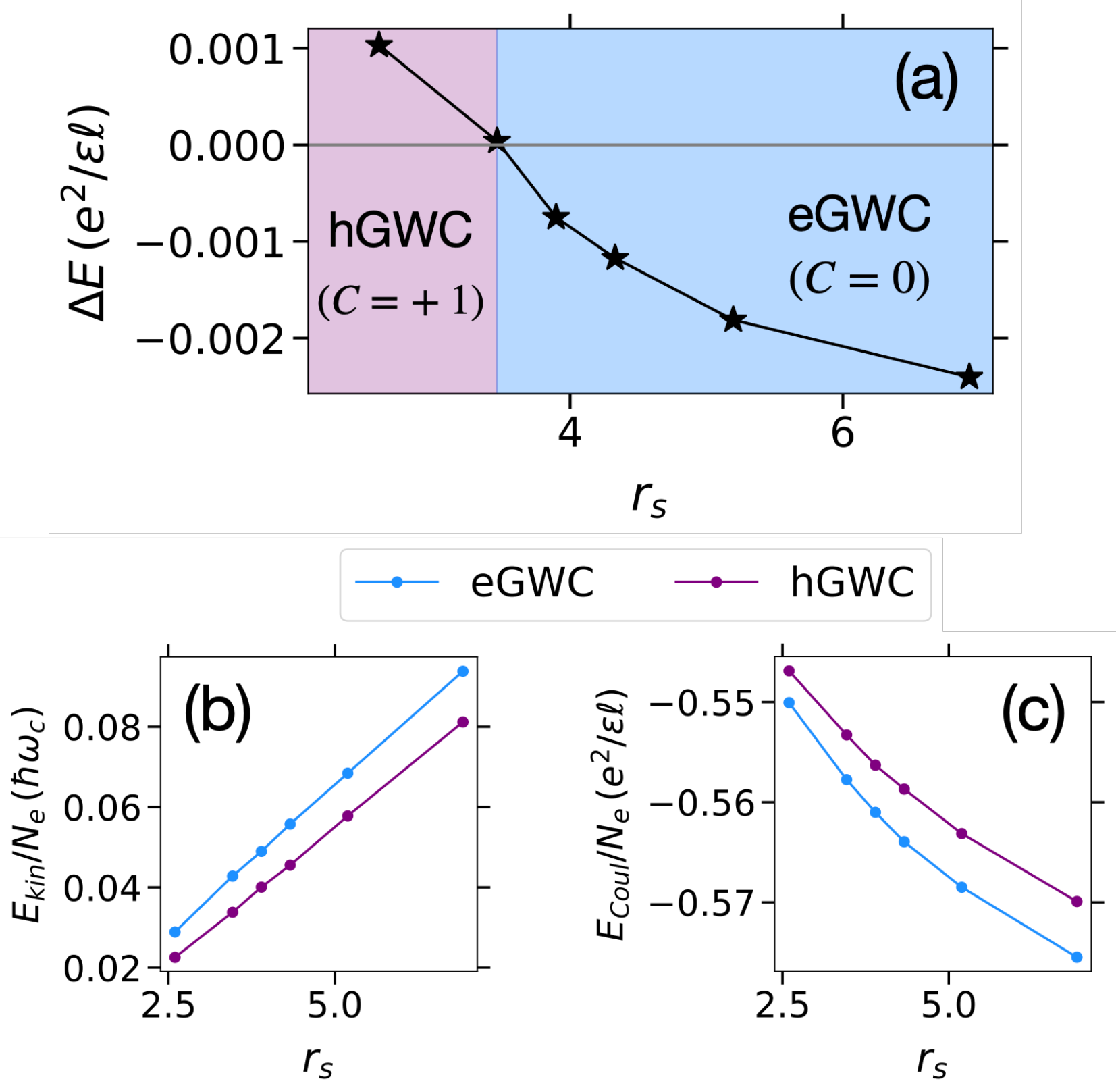}
    \caption{Competition between the electron GWC and the hole GWC as a function of $r_s$, for fixed $B_1=+0.75$. (a) Difference between the energy per particle of the electron GWC and the hole GWC, as a function of $r_s$. Band mixing drives a topological phase transition from the hole crystal to the electron crystal. (b) Kinetic and (c) Coulomb energies per particle for both crystals as a function of $r_s$. Band mixing favors electronic orbital localization, lowering the Coulomb energy of the electron crystal state, which becomes the ground state for large enough $r_s$.}
    \label{fig:eWC_vs_hWC}
\end{figure}

Fig.~\ref{fig:eWC_vs_hWC}(a) shows the difference between the electron and hole GWC energies per particle as a function of $r_s$, while Fig.~\ref{fig:eWC_vs_hWC}(b),(c) show the kinetic and Coulomb contributions to the energies. Band mixing allows electronic orbitals to localize further, lowering the Coulomb energy of the electron GWC at the expense of an increased kinetic energy. In contrast, the Coulomb energy gain of the hole GWC due to band mixing is lower, so strong band mixing favors the electron GWC despite its incommensurability with the magnetic field.

The hole GWC survives only up to a finite value of $B_1$, even at the weakest band mixing considered in our calculations, beyond which the electron GWC becomes the ground state. This behavior can be understood from the charge density of the fully filled AC band. In the limit $B_1\gg1$, the charge density is concentrated at the AB/BA sites and suppressed at the AA sites, which limits the localization of holes at AA sites. The resulting increase in Coulomb energy due to more extended hole orbitals makes the hole GWC energetically unfavorable for large $B_1>0$.

In the regime $B_1<0$, our calculations show that once the FCI becomes unstable, a honeycomb electron GWC is the ground state. Other possible ground state candidates are an incommensurate electron GWC and a commensurate hole GWC. We find neither of those to be stable in this regime. A recent band-projected ED study \cite{Guerci_AC_SC}, which uses a Haldane pseudopotential interaction, suggests that the ground state for $B_1<0$ could be a superconductor. For the Coulomb interaction we find no evidence of superconductivity within the parameter range studied here. Determining how band mixing and interaction screening affect the stability of a possible superconducting phase in the AC model is an important question for guiding future experiments, but is beyond the scope of this work. 

Finally, at $B_1=0$ we estimate the freezing of the Jain state to happen at $r_s\approx 44.5$, or equivalently $\kappa\approx25.7$. To our knowledge, the critical value for the freezing of the Jain state at $\nu=2/3$ has not been reported before. 

\section{Summary and Discussion}
\label{sec:Discussion}
We have introduced a physics-informed neural-network-based variational Monte Carlo approach for continuum models, which combines the expressivity of NQS with physical insight built into the wave function. Our approach accurately represents both topologically ordered and crystalline states, includes band mixing, and can access system sizes beyond the reach of currently available high-accuracy many-body approaches. We have used our method to determine the phase diagram of the AC model at fractional fillings of the lowest band, providing new insight into how discrete translational symmetry and band mixing affect fractionalized states and their competition with charge-ordered phases. These questions are directly relevant to moiré materials such as tMoTe$_2$ and rhombohedral graphene.

We find that the Laughlin-like state at $\nu=1/3$ and the Jain-like state at $\nu=2/3$ respond differently to discrete translation symmetry. In particular, as quantum geometry becomes more nonuniform, the energy of the former remains nearly constant, while the energy of the latter decreases. Both FCI phases are destabilized in favor of crystalline order 
for sufficiently strong interactions or sufficiently strong nonuniform Berry curvature. In the limit of uniform field ($B_1=0$), we obtain that the critical values for the FQH-to-crystal transitions are: $\kappa_c\approx 23$ at $\nu=1/3$ and $\kappa_c\approx 26$ at $\nu=2/3$.

We find that the interplay between interactions and the Berry curvature distribution determines the structure and Chern number of the crystalline ground state. Remarkably, phases commensurate with the periodicity of the Hamiltonian are not always favored: we find an incommensurate electron generalized Wigner crystal ground state at $\nu=2/3$. Our results also show that strong band mixing favors electron Wigner crystals over hole Wigner crystals, by enabling further electronic orbital localization, which lowers the Coulomb energy of the electron crystal. Band mixing can also stabilize phases absent when interactions are projected to the lowest band, such as the $C_3$-crystal at $\nu=1/3$.

Because our framework enables access to larger system sizes than other existing approaches, it can be used to study filling fractions that were previously out of reach. Specifically, we can study the ground states of slightly doped FCIs, where a dilute gas of anyons with a non-trivial dispersion emerges \cite{Goncalves2026Spinless,Yan2026Anyon,Schleith2026Anyon,iyer2026dispersion}. We can also study distinct filling fractions within $2/3<\nu<1$, where recent band-projected studies \cite{Guerci_AC_SC,ouyang2026chiral,Xu2025Chiral,wang2025chiral} predict topological superconductivity in the AC Hamiltonian and related models. Additionally, the valley degree of freedom can be introduced into our method to investigate the stability of fractional topological insulators at $\nu=4/3$ \cite{Levin2009Fractional,Kwan2026Regarding,Wang2026Candidate}. We leave these directions for future work.

Finally, we comment on the connection of our results to semiconductor moiré materials such as tMoTe$_2$, which combine moiré discrete translational symmetry with nonuniform Berry curvature, and have a relatively flat topmost valence band. In such systems, comparisons between band-projected numerics and experiment indicate band mixing plays an important role \cite{Yu2024Fractional,Abouelkomsan2024Band}. Our framework provides a scalable approach to incorporating multiband effects in continuum models of moiré materials. The hole generalized Wigner crystal phase that we find at $\nu=2/3$ provides a candidate for the re-entrant integer quantum anomalous Hall state observed close to this filling \cite{xu2025superconductivity}. More broadly, the contrasting phase diagrams at $\nu=1/3$ and $\nu=2/3$ offer insight into the phase competition that may give rise to the experimentally observed asymmetry between these fillings.\\\\

{\it Acknowledgments -- }We thank Jennifer Cano, Nicolas Regnault and Tomohiro Soejima for helpful discussions. The Flatiron Institute is a division of the Simons Foundation. This work was supported in part by the NSF MRSEC program through Columbia in the Center for Advanced Electronic Materials (DMR-2614607). This research was supported in part by grant NSF PHY-2309135 to the Kavli Institute for Theoretical Physics (KITP).

\bibliography{references}

\begin{thebibliography}{83}%
\makeatletter
\providecommand \@ifxundefined [1]{%
 \@ifx{#1\undefined}
}%
\providecommand \@ifnum [1]{%
 \ifnum #1\expandafter \@firstoftwo
 \else \expandafter \@secondoftwo
 \fi
}%
\providecommand \@ifx [1]{%
 \ifx #1\expandafter \@firstoftwo
 \else \expandafter \@secondoftwo
 \fi
}%
\providecommand \natexlab [1]{#1}%
\providecommand \enquote  [1]{``#1''}%
\providecommand \bibnamefont  [1]{#1}%
\providecommand \bibfnamefont [1]{#1}%
\providecommand \citenamefont [1]{#1}%
\providecommand \href@noop [0]{\@secondoftwo}%
\providecommand \href [0]{\begingroup \@sanitize@url \@href}%
\providecommand \@href[1]{\@@startlink{#1}\@@href}%
\providecommand \@@href[1]{\endgroup#1\@@endlink}%
\providecommand \@sanitize@url [0]{\catcode `\\12\catcode `\$12\catcode
  `\&12\catcode `\#12\catcode `\^12\catcode `\_12\catcode `\%12\relax}%
\providecommand \@@startlink[1]{}%
\providecommand \@@endlink[0]{}%
\providecommand \url  [0]{\begingroup\@sanitize@url \@url }%
\providecommand \@url [1]{\endgroup\@href {#1}{\urlprefix }}%
\providecommand \urlprefix  [0]{URL }%
\providecommand \Eprint [0]{\href }%
\providecommand \doibase [0]{https://doi.org/}%
\providecommand \selectlanguage [0]{\@gobble}%
\providecommand \bibinfo  [0]{\@secondoftwo}%
\providecommand \bibfield  [0]{\@secondoftwo}%
\providecommand \translation [1]{[#1]}%
\providecommand \BibitemOpen [0]{}%
\providecommand \bibitemStop [0]{}%
\providecommand \bibitemNoStop [0]{.\EOS\space}%
\providecommand \EOS [0]{\spacefactor3000\relax}%
\providecommand \BibitemShut  [1]{\csname bibitem#1\endcsname}%
\let\auto@bib@innerbib\@empty
\bibitem [{\citenamefont {Cai}\ \emph {et~al.}(2023)\citenamefont {Cai},
  \citenamefont {Anderson}, \citenamefont {Wang}, \citenamefont {Zhang},
  \citenamefont {Liu}, \citenamefont {Holtzmann}, \citenamefont {Zhang},
  \citenamefont {Fan}, \citenamefont {Taniguchi}, \citenamefont {Watanabe},
  \citenamefont {Ran}, \citenamefont {Cao}, \citenamefont {Fu}, \citenamefont
  {Xiao}, \citenamefont {Yao},\ and\ \citenamefont {Xu}}]{cai2023signatures}%
  \BibitemOpen
  \bibfield  {author} {\bibinfo {author} {\bibfnamefont {J.}~\bibnamefont
  {Cai}}, \bibinfo {author} {\bibfnamefont {E.}~\bibnamefont {Anderson}},
  \bibinfo {author} {\bibfnamefont {C.}~\bibnamefont {Wang}}, \bibinfo {author}
  {\bibfnamefont {X.}~\bibnamefont {Zhang}}, \bibinfo {author} {\bibfnamefont
  {X.}~\bibnamefont {Liu}}, \bibinfo {author} {\bibfnamefont {W.}~\bibnamefont
  {Holtzmann}}, \bibinfo {author} {\bibfnamefont {Y.}~\bibnamefont {Zhang}},
  \bibinfo {author} {\bibfnamefont {F.}~\bibnamefont {Fan}}, \bibinfo {author}
  {\bibfnamefont {T.}~\bibnamefont {Taniguchi}}, \bibinfo {author}
  {\bibfnamefont {K.}~\bibnamefont {Watanabe}}, \bibinfo {author}
  {\bibfnamefont {Y.}~\bibnamefont {Ran}}, \bibinfo {author} {\bibfnamefont
  {T.}~\bibnamefont {Cao}}, \bibinfo {author} {\bibfnamefont {L.}~\bibnamefont
  {Fu}}, \bibinfo {author} {\bibfnamefont {D.}~\bibnamefont {Xiao}}, \bibinfo
  {author} {\bibfnamefont {W.}~\bibnamefont {Yao}},\ and\ \bibinfo {author}
  {\bibfnamefont {X.}~\bibnamefont {Xu}},\ }\bibfield  {title} {\bibinfo
  {title} {Signatures of fractional quantum anomalous {H}all states in twisted
  {M}o{T}e$_2$},\ }\href {https://doi.org/10.1038/s41586-023-06289-w}
  {\bibfield  {journal} {\bibinfo  {journal} {Nature}\ }\textbf {\bibinfo
  {volume} {622}},\ \bibinfo {pages} {63} (\bibinfo {year} {2023})}\BibitemShut
  {NoStop}%
\bibitem [{\citenamefont {Zeng}\ \emph {et~al.}(2023)\citenamefont {Zeng},
  \citenamefont {Xia}, \citenamefont {Kang}, \citenamefont {Zhu}, \citenamefont
  {Kn{\"u}ppel}, \citenamefont {Vaswani}, \citenamefont {Watanabe},
  \citenamefont {Taniguchi}, \citenamefont {Mak},\ and\ \citenamefont
  {Shan}}]{zeng2023thermodynamic}%
  \BibitemOpen
  \bibfield  {author} {\bibinfo {author} {\bibfnamefont {Y.}~\bibnamefont
  {Zeng}}, \bibinfo {author} {\bibfnamefont {Z.}~\bibnamefont {Xia}}, \bibinfo
  {author} {\bibfnamefont {K.}~\bibnamefont {Kang}}, \bibinfo {author}
  {\bibfnamefont {J.}~\bibnamefont {Zhu}}, \bibinfo {author} {\bibfnamefont
  {P.}~\bibnamefont {Kn{\"u}ppel}}, \bibinfo {author} {\bibfnamefont
  {C.}~\bibnamefont {Vaswani}}, \bibinfo {author} {\bibfnamefont
  {K.}~\bibnamefont {Watanabe}}, \bibinfo {author} {\bibfnamefont
  {T.}~\bibnamefont {Taniguchi}}, \bibinfo {author} {\bibfnamefont {K.~F.}\
  \bibnamefont {Mak}},\ and\ \bibinfo {author} {\bibfnamefont {J.}~\bibnamefont
  {Shan}},\ }\bibfield  {title} {\bibinfo {title} {Thermodynamic evidence of
  fractional {C}hern insulator in moir{\'e} {M}o{T}e$_2$},\ }\href
  {https://doi.org/10.1038/s41586-023-06452-3} {\bibfield  {journal} {\bibinfo
  {journal} {Nature}\ }\textbf {\bibinfo {volume} {622}},\ \bibinfo {pages}
  {69} (\bibinfo {year} {2023})}\BibitemShut {NoStop}%
\bibitem [{\citenamefont {Park}\ \emph {et~al.}(2023)\citenamefont {Park},
  \citenamefont {Cai}, \citenamefont {Anderson}, \citenamefont {Zhang},
  \citenamefont {Zhu}, \citenamefont {Liu}, \citenamefont {Wang}, \citenamefont
  {Holtzmann}, \citenamefont {Hu}, \citenamefont {Liu}, \citenamefont
  {Taniguchi}, \citenamefont {Watanabe}, \citenamefont {Chu}, \citenamefont
  {Cao}, \citenamefont {Fu}, \citenamefont {Yao}, \citenamefont {Chang},
  \citenamefont {Cobden}, \citenamefont {Xiao},\ and\ \citenamefont
  {Xu}}]{park2023observation}%
  \BibitemOpen
  \bibfield  {author} {\bibinfo {author} {\bibfnamefont {H.}~\bibnamefont
  {Park}}, \bibinfo {author} {\bibfnamefont {J.}~\bibnamefont {Cai}}, \bibinfo
  {author} {\bibfnamefont {E.}~\bibnamefont {Anderson}}, \bibinfo {author}
  {\bibfnamefont {Y.}~\bibnamefont {Zhang}}, \bibinfo {author} {\bibfnamefont
  {J.}~\bibnamefont {Zhu}}, \bibinfo {author} {\bibfnamefont {X.}~\bibnamefont
  {Liu}}, \bibinfo {author} {\bibfnamefont {C.}~\bibnamefont {Wang}}, \bibinfo
  {author} {\bibfnamefont {W.}~\bibnamefont {Holtzmann}}, \bibinfo {author}
  {\bibfnamefont {C.}~\bibnamefont {Hu}}, \bibinfo {author} {\bibfnamefont
  {Z.}~\bibnamefont {Liu}}, \bibinfo {author} {\bibfnamefont {T.}~\bibnamefont
  {Taniguchi}}, \bibinfo {author} {\bibfnamefont {K.}~\bibnamefont {Watanabe}},
  \bibinfo {author} {\bibfnamefont {J.-H.}\ \bibnamefont {Chu}}, \bibinfo
  {author} {\bibfnamefont {T.}~\bibnamefont {Cao}}, \bibinfo {author}
  {\bibfnamefont {L.}~\bibnamefont {Fu}}, \bibinfo {author} {\bibfnamefont
  {W.}~\bibnamefont {Yao}}, \bibinfo {author} {\bibfnamefont {C.-Z.}\
  \bibnamefont {Chang}}, \bibinfo {author} {\bibfnamefont {D.}~\bibnamefont
  {Cobden}}, \bibinfo {author} {\bibfnamefont {D.}~\bibnamefont {Xiao}},\ and\
  \bibinfo {author} {\bibfnamefont {X.}~\bibnamefont {Xu}},\ }\bibfield
  {title} {\bibinfo {title} {Observation of fractionally quantized anomalous
  {H}all effect},\ }\href {https://doi.org/10.1038/s41586-023-06536-0}
  {\bibfield  {journal} {\bibinfo  {journal} {Nature}\ }\textbf {\bibinfo
  {volume} {622}},\ \bibinfo {pages} {74} (\bibinfo {year} {2023})}\BibitemShut
  {NoStop}%
\bibitem [{\citenamefont {Xu}\ \emph {et~al.}(2023)\citenamefont {Xu},
  \citenamefont {Sun}, \citenamefont {Jia}, \citenamefont {Liu}, \citenamefont
  {Xu}, \citenamefont {Li}, \citenamefont {Gu}, \citenamefont {Watanabe},
  \citenamefont {Taniguchi}, \citenamefont {Tong}, \citenamefont {Jia},
  \citenamefont {Shi}, \citenamefont {Jiang}, \citenamefont {Zhang},
  \citenamefont {Liu},\ and\ \citenamefont {Li}}]{xu2023observation}%
  \BibitemOpen
  \bibfield  {author} {\bibinfo {author} {\bibfnamefont {F.}~\bibnamefont
  {Xu}}, \bibinfo {author} {\bibfnamefont {Z.}~\bibnamefont {Sun}}, \bibinfo
  {author} {\bibfnamefont {T.}~\bibnamefont {Jia}}, \bibinfo {author}
  {\bibfnamefont {C.}~\bibnamefont {Liu}}, \bibinfo {author} {\bibfnamefont
  {C.}~\bibnamefont {Xu}}, \bibinfo {author} {\bibfnamefont {C.}~\bibnamefont
  {Li}}, \bibinfo {author} {\bibfnamefont {Y.}~\bibnamefont {Gu}}, \bibinfo
  {author} {\bibfnamefont {K.}~\bibnamefont {Watanabe}}, \bibinfo {author}
  {\bibfnamefont {T.}~\bibnamefont {Taniguchi}}, \bibinfo {author}
  {\bibfnamefont {B.}~\bibnamefont {Tong}}, \bibinfo {author} {\bibfnamefont
  {J.}~\bibnamefont {Jia}}, \bibinfo {author} {\bibfnamefont {Z.}~\bibnamefont
  {Shi}}, \bibinfo {author} {\bibfnamefont {S.}~\bibnamefont {Jiang}}, \bibinfo
  {author} {\bibfnamefont {Y.}~\bibnamefont {Zhang}}, \bibinfo {author}
  {\bibfnamefont {X.}~\bibnamefont {Liu}},\ and\ \bibinfo {author}
  {\bibfnamefont {T.}~\bibnamefont {Li}},\ }\bibfield  {title} {\bibinfo
  {title} {Observation of integer and fractional quantum anomalous {H}all
  effects in twisted bilayer {M}o{T}e$_2$},\ }\href
  {https://doi.org/10.1103/PhysRevX.13.031037} {\bibfield  {journal} {\bibinfo
  {journal} {Phys. Rev. X}\ }\textbf {\bibinfo {volume} {13}},\ \bibinfo
  {pages} {031037} (\bibinfo {year} {2023})}\BibitemShut {NoStop}%
\bibitem [{\citenamefont {Lu}\ \emph {et~al.}(2024)\citenamefont {Lu},
  \citenamefont {Han}, \citenamefont {Yao}, \citenamefont {Reddy},
  \citenamefont {Yang}, \citenamefont {Seo}, \citenamefont {Watanabe},
  \citenamefont {Taniguchi}, \citenamefont {Fu},\ and\ \citenamefont
  {Ju}}]{lu2024fractional}%
  \BibitemOpen
  \bibfield  {author} {\bibinfo {author} {\bibfnamefont {Z.}~\bibnamefont
  {Lu}}, \bibinfo {author} {\bibfnamefont {T.}~\bibnamefont {Han}}, \bibinfo
  {author} {\bibfnamefont {Y.}~\bibnamefont {Yao}}, \bibinfo {author}
  {\bibfnamefont {A.~P.}\ \bibnamefont {Reddy}}, \bibinfo {author}
  {\bibfnamefont {J.}~\bibnamefont {Yang}}, \bibinfo {author} {\bibfnamefont
  {J.}~\bibnamefont {Seo}}, \bibinfo {author} {\bibfnamefont {K.}~\bibnamefont
  {Watanabe}}, \bibinfo {author} {\bibfnamefont {T.}~\bibnamefont {Taniguchi}},
  \bibinfo {author} {\bibfnamefont {L.}~\bibnamefont {Fu}},\ and\ \bibinfo
  {author} {\bibfnamefont {L.}~\bibnamefont {Ju}},\ }\bibfield  {title}
  {\bibinfo {title} {Fractional quantum anomalous {H}all effect in multilayer
  graphene},\ }\href {https://doi.org/10.1038/s41586-023-07010-7} {\bibfield
  {journal} {\bibinfo  {journal} {Nature}\ }\textbf {\bibinfo {volume} {626}},\
  \bibinfo {pages} {759} (\bibinfo {year} {2024})}\BibitemShut {NoStop}%
\bibitem [{\citenamefont {Sun}\ \emph {et~al.}(2026)\citenamefont {Sun},
  \citenamefont {Xu}, \citenamefont {Li}, \citenamefont {Jiang}, \citenamefont
  {Gao}, \citenamefont {Xu}, \citenamefont {Jia}, \citenamefont {Cheng},
  \citenamefont {Zhang}, \citenamefont {Tian}, \citenamefont {Watanabe},
  \citenamefont {Taniguchi}, \citenamefont {Jia}, \citenamefont {Jiang},
  \citenamefont {Zhang}, \citenamefont {Zhang}, \citenamefont {Lei},
  \citenamefont {Liu},\ and\ \citenamefont {Li}}]{sun2026twistangle}%
  \BibitemOpen
  \bibfield  {author} {\bibinfo {author} {\bibfnamefont {Z.}~\bibnamefont
  {Sun}}, \bibinfo {author} {\bibfnamefont {F.}~\bibnamefont {Xu}}, \bibinfo
  {author} {\bibfnamefont {J.}~\bibnamefont {Li}}, \bibinfo {author}
  {\bibfnamefont {Y.}~\bibnamefont {Jiang}}, \bibinfo {author} {\bibfnamefont
  {J.}~\bibnamefont {Gao}}, \bibinfo {author} {\bibfnamefont {C.}~\bibnamefont
  {Xu}}, \bibinfo {author} {\bibfnamefont {T.}~\bibnamefont {Jia}}, \bibinfo
  {author} {\bibfnamefont {K.}~\bibnamefont {Cheng}}, \bibinfo {author}
  {\bibfnamefont {J.}~\bibnamefont {Zhang}}, \bibinfo {author} {\bibfnamefont
  {W.}~\bibnamefont {Tian}}, \bibinfo {author} {\bibfnamefont {K.}~\bibnamefont
  {Watanabe}}, \bibinfo {author} {\bibfnamefont {T.}~\bibnamefont {Taniguchi}},
  \bibinfo {author} {\bibfnamefont {J.}~\bibnamefont {Jia}}, \bibinfo {author}
  {\bibfnamefont {S.}~\bibnamefont {Jiang}}, \bibinfo {author} {\bibfnamefont
  {Y.}~\bibnamefont {Zhang}}, \bibinfo {author} {\bibfnamefont
  {Y.}~\bibnamefont {Zhang}}, \bibinfo {author} {\bibfnamefont
  {S.}~\bibnamefont {Lei}}, \bibinfo {author} {\bibfnamefont {X.}~\bibnamefont
  {Liu}},\ and\ \bibinfo {author} {\bibfnamefont {T.}~\bibnamefont {Li}},\
  }\href {https://arxiv.org/abs/2603.16412} {\bibinfo {title} {Twist-angle
  evolution from valley-polarized fractional topological phases to
  valley-degenerate superconductivity in twisted bilayer {M}o{T}e$_2$}}
  (\bibinfo {year} {2026}),\ \Eprint {https://arxiv.org/abs/2603.16412}
  {arXiv:2603.16412 [cond-mat.mes-hall]} \BibitemShut {NoStop}%
\bibitem [{\citenamefont {Lu}\ \emph {et~al.}(2025)\citenamefont {Lu},
  \citenamefont {Han}, \citenamefont {Yao}, \citenamefont {Hadjri},
  \citenamefont {Yang}, \citenamefont {Seo}, \citenamefont {Shi}, \citenamefont
  {Ye}, \citenamefont {Watanabe}, \citenamefont {Taniguchi},\ and\
  \citenamefont {Ju}}]{Lu2025Extended}%
  \BibitemOpen
  \bibfield  {author} {\bibinfo {author} {\bibfnamefont {Z.}~\bibnamefont
  {Lu}}, \bibinfo {author} {\bibfnamefont {T.}~\bibnamefont {Han}}, \bibinfo
  {author} {\bibfnamefont {Y.}~\bibnamefont {Yao}}, \bibinfo {author}
  {\bibfnamefont {Z.}~\bibnamefont {Hadjri}}, \bibinfo {author} {\bibfnamefont
  {J.}~\bibnamefont {Yang}}, \bibinfo {author} {\bibfnamefont {J.}~\bibnamefont
  {Seo}}, \bibinfo {author} {\bibfnamefont {L.}~\bibnamefont {Shi}}, \bibinfo
  {author} {\bibfnamefont {S.}~\bibnamefont {Ye}}, \bibinfo {author}
  {\bibfnamefont {K.}~\bibnamefont {Watanabe}}, \bibinfo {author}
  {\bibfnamefont {T.}~\bibnamefont {Taniguchi}},\ and\ \bibinfo {author}
  {\bibfnamefont {L.}~\bibnamefont {Ju}},\ }\bibfield  {title} {\bibinfo
  {title} {Extended quantum anomalous {H}all states in graphene/h{BN} moir{\'e}
  superlattices},\ }\href {https://doi.org/10.1038/s41586-024-08470-1}
  {\bibfield  {journal} {\bibinfo  {journal} {Nature}\ }\textbf {\bibinfo
  {volume} {637}},\ \bibinfo {pages} {1090} (\bibinfo {year}
  {2025})}\BibitemShut {NoStop}%
\bibitem [{\citenamefont {Aronson}\ \emph {et~al.}(2025)\citenamefont
  {Aronson}, \citenamefont {Han}, \citenamefont {Lu}, \citenamefont {Yao},
  \citenamefont {Butler}, \citenamefont {Watanabe}, \citenamefont {Taniguchi},
  \citenamefont {Ju},\ and\ \citenamefont {Ashoori}}]{Aronson2025Displacement}%
  \BibitemOpen
  \bibfield  {author} {\bibinfo {author} {\bibfnamefont {S.~H.}\ \bibnamefont
  {Aronson}}, \bibinfo {author} {\bibfnamefont {T.}~\bibnamefont {Han}},
  \bibinfo {author} {\bibfnamefont {Z.}~\bibnamefont {Lu}}, \bibinfo {author}
  {\bibfnamefont {Y.}~\bibnamefont {Yao}}, \bibinfo {author} {\bibfnamefont
  {J.~P.}\ \bibnamefont {Butler}}, \bibinfo {author} {\bibfnamefont
  {K.}~\bibnamefont {Watanabe}}, \bibinfo {author} {\bibfnamefont
  {T.}~\bibnamefont {Taniguchi}}, \bibinfo {author} {\bibfnamefont
  {L.}~\bibnamefont {Ju}},\ and\ \bibinfo {author} {\bibfnamefont {R.~C.}\
  \bibnamefont {Ashoori}},\ }\bibfield  {title} {\bibinfo {title} {Displacement
  field-controlled fractional {C}hern insulators and charge density waves in a
  graphene/h{BN} moir\'e superlattice},\ }\href
  {https://doi.org/10.1103/75gl-jzl6} {\bibfield  {journal} {\bibinfo
  {journal} {Phys. Rev. X}\ }\textbf {\bibinfo {volume} {15}},\ \bibinfo
  {pages} {031026} (\bibinfo {year} {2025})}\BibitemShut {NoStop}%
\bibitem [{\citenamefont {Xu}\ \emph {et~al.}(2025{\natexlab{a}})\citenamefont
  {Xu}, \citenamefont {Sun}, \citenamefont {Li}, \citenamefont {Zheng},
  \citenamefont {Xu}, \citenamefont {Gao}, \citenamefont {Jia}, \citenamefont
  {Watanabe}, \citenamefont {Taniguchi}, \citenamefont {Tong}, \citenamefont
  {Lu}, \citenamefont {Jia}, \citenamefont {Shi}, \citenamefont {Jiang},
  \citenamefont {Zhang}, \citenamefont {Zhang}, \citenamefont {Lei},
  \citenamefont {Liu},\ and\ \citenamefont {Li}}]{xu2025superconductivity}%
  \BibitemOpen
  \bibfield  {author} {\bibinfo {author} {\bibfnamefont {F.}~\bibnamefont
  {Xu}}, \bibinfo {author} {\bibfnamefont {Z.}~\bibnamefont {Sun}}, \bibinfo
  {author} {\bibfnamefont {J.}~\bibnamefont {Li}}, \bibinfo {author}
  {\bibfnamefont {C.}~\bibnamefont {Zheng}}, \bibinfo {author} {\bibfnamefont
  {C.}~\bibnamefont {Xu}}, \bibinfo {author} {\bibfnamefont {J.}~\bibnamefont
  {Gao}}, \bibinfo {author} {\bibfnamefont {T.}~\bibnamefont {Jia}}, \bibinfo
  {author} {\bibfnamefont {K.}~\bibnamefont {Watanabe}}, \bibinfo {author}
  {\bibfnamefont {T.}~\bibnamefont {Taniguchi}}, \bibinfo {author}
  {\bibfnamefont {B.}~\bibnamefont {Tong}}, \bibinfo {author} {\bibfnamefont
  {L.}~\bibnamefont {Lu}}, \bibinfo {author} {\bibfnamefont {J.}~\bibnamefont
  {Jia}}, \bibinfo {author} {\bibfnamefont {Z.}~\bibnamefont {Shi}}, \bibinfo
  {author} {\bibfnamefont {S.}~\bibnamefont {Jiang}}, \bibinfo {author}
  {\bibfnamefont {Y.}~\bibnamefont {Zhang}}, \bibinfo {author} {\bibfnamefont
  {Y.}~\bibnamefont {Zhang}}, \bibinfo {author} {\bibfnamefont
  {S.}~\bibnamefont {Lei}}, \bibinfo {author} {\bibfnamefont {X.}~\bibnamefont
  {Liu}},\ and\ \bibinfo {author} {\bibfnamefont {T.}~\bibnamefont {Li}},\
  }\href {https://arxiv.org/abs/2504.06972} {\bibinfo {title} {Signatures of
  unconventional superconductivity near reentrant and fractional quantum
  anomalous {H}all insulators}} (\bibinfo {year} {2025}{\natexlab{a}}),\
  \Eprint {https://arxiv.org/abs/2504.06972} {arXiv:2504.06972
  [cond-mat.mes-hall]} \BibitemShut {NoStop}%
\bibitem [{\citenamefont {Han}\ \emph {et~al.}(2025)\citenamefont {Han},
  \citenamefont {Lu}, \citenamefont {Hadjri}, \citenamefont {Shi},
  \citenamefont {Wu}, \citenamefont {Xu}, \citenamefont {Yao}, \citenamefont
  {Cotten}, \citenamefont {Sharifi~Sedeh}, \citenamefont {Weldeyesus},
  \citenamefont {Yang}, \citenamefont {Seo}, \citenamefont {Ye}, \citenamefont
  {Zhou}, \citenamefont {Liu}, \citenamefont {Shi}, \citenamefont {Hua},
  \citenamefont {Watanabe}, \citenamefont {Taniguchi}, \citenamefont {Xiong},
  \citenamefont {Zumb{\"u}hl}, \citenamefont {Fu},\ and\ \citenamefont
  {Ju}}]{Han2025Signatures}%
  \BibitemOpen
  \bibfield  {author} {\bibinfo {author} {\bibfnamefont {T.}~\bibnamefont
  {Han}}, \bibinfo {author} {\bibfnamefont {Z.}~\bibnamefont {Lu}}, \bibinfo
  {author} {\bibfnamefont {Z.}~\bibnamefont {Hadjri}}, \bibinfo {author}
  {\bibfnamefont {L.}~\bibnamefont {Shi}}, \bibinfo {author} {\bibfnamefont
  {Z.}~\bibnamefont {Wu}}, \bibinfo {author} {\bibfnamefont {W.}~\bibnamefont
  {Xu}}, \bibinfo {author} {\bibfnamefont {Y.}~\bibnamefont {Yao}}, \bibinfo
  {author} {\bibfnamefont {A.~A.}\ \bibnamefont {Cotten}}, \bibinfo {author}
  {\bibfnamefont {O.}~\bibnamefont {Sharifi~Sedeh}}, \bibinfo {author}
  {\bibfnamefont {H.}~\bibnamefont {Weldeyesus}}, \bibinfo {author}
  {\bibfnamefont {J.}~\bibnamefont {Yang}}, \bibinfo {author} {\bibfnamefont
  {J.}~\bibnamefont {Seo}}, \bibinfo {author} {\bibfnamefont {S.}~\bibnamefont
  {Ye}}, \bibinfo {author} {\bibfnamefont {M.}~\bibnamefont {Zhou}}, \bibinfo
  {author} {\bibfnamefont {H.}~\bibnamefont {Liu}}, \bibinfo {author}
  {\bibfnamefont {G.}~\bibnamefont {Shi}}, \bibinfo {author} {\bibfnamefont
  {Z.}~\bibnamefont {Hua}}, \bibinfo {author} {\bibfnamefont {K.}~\bibnamefont
  {Watanabe}}, \bibinfo {author} {\bibfnamefont {T.}~\bibnamefont {Taniguchi}},
  \bibinfo {author} {\bibfnamefont {P.}~\bibnamefont {Xiong}}, \bibinfo
  {author} {\bibfnamefont {D.~M.}\ \bibnamefont {Zumb{\"u}hl}}, \bibinfo
  {author} {\bibfnamefont {L.}~\bibnamefont {Fu}},\ and\ \bibinfo {author}
  {\bibfnamefont {L.}~\bibnamefont {Ju}},\ }\bibfield  {title} {\bibinfo
  {title} {Signatures of chiral superconductivity in rhombohedral graphene},\
  }\href@noop {} {\bibfield  {journal} {\bibinfo  {journal} {Nature}\ }\textbf
  {\bibinfo {volume} {643}},\ \bibinfo {pages} {654} (\bibinfo {year}
  {2025})}\BibitemShut {NoStop}%
\bibitem [{\citenamefont {Wang}\ \emph {et~al.}(2021)\citenamefont {Wang},
  \citenamefont {Cano}, \citenamefont {Millis}, \citenamefont {Liu},\ and\
  \citenamefont {Yang}}]{wang2021exact}%
  \BibitemOpen
  \bibfield  {author} {\bibinfo {author} {\bibfnamefont {J.}~\bibnamefont
  {Wang}}, \bibinfo {author} {\bibfnamefont {J.}~\bibnamefont {Cano}}, \bibinfo
  {author} {\bibfnamefont {A.~J.}\ \bibnamefont {Millis}}, \bibinfo {author}
  {\bibfnamefont {Z.}~\bibnamefont {Liu}},\ and\ \bibinfo {author}
  {\bibfnamefont {B.}~\bibnamefont {Yang}},\ }\bibfield  {title} {\bibinfo
  {title} {Exact {L}andau level description of geometry and interaction in a
  flatband},\ }\href {https://doi.org/10.1103/PhysRevLett.127.246403}
  {\bibfield  {journal} {\bibinfo  {journal} {Phys. Rev. Lett.}\ }\textbf
  {\bibinfo {volume} {127}},\ \bibinfo {pages} {246403} (\bibinfo {year}
  {2021})}\BibitemShut {NoStop}%
\bibitem [{\citenamefont {Ledwith}\ \emph {et~al.}(2020)\citenamefont
  {Ledwith}, \citenamefont {Tarnopolsky}, \citenamefont {Khalaf},\ and\
  \citenamefont {Vishwanath}}]{ledwith2020fractional}%
  \BibitemOpen
  \bibfield  {author} {\bibinfo {author} {\bibfnamefont {P.~J.}\ \bibnamefont
  {Ledwith}}, \bibinfo {author} {\bibfnamefont {G.}~\bibnamefont
  {Tarnopolsky}}, \bibinfo {author} {\bibfnamefont {E.}~\bibnamefont
  {Khalaf}},\ and\ \bibinfo {author} {\bibfnamefont {A.}~\bibnamefont
  {Vishwanath}},\ }\bibfield  {title} {\bibinfo {title} {Fractional {C}hern
  insulator states in twisted bilayer graphene: An analytical approach},\
  }\href {https://doi.org/10.1103/PhysRevResearch.2.023237} {\bibfield
  {journal} {\bibinfo  {journal} {Phys. Rev. Research}\ }\textbf {\bibinfo
  {volume} {2}},\ \bibinfo {pages} {023237} (\bibinfo {year}
  {2020})}\BibitemShut {NoStop}%
\bibitem [{\citenamefont {Ledwith}\ \emph {et~al.}(2023)\citenamefont
  {Ledwith}, \citenamefont {Vishwanath},\ and\ \citenamefont
  {Parker}}]{ledwith2023vortexability}%
  \BibitemOpen
  \bibfield  {author} {\bibinfo {author} {\bibfnamefont {P.~J.}\ \bibnamefont
  {Ledwith}}, \bibinfo {author} {\bibfnamefont {A.}~\bibnamefont
  {Vishwanath}},\ and\ \bibinfo {author} {\bibfnamefont {D.~E.}\ \bibnamefont
  {Parker}},\ }\bibfield  {title} {\bibinfo {title} {Vortexability: A unifying
  criterion for ideal fractional {C}hern insulators},\ }\href
  {https://doi.org/10.1103/PhysRevB.108.205144} {\bibfield  {journal} {\bibinfo
   {journal} {Phys. Rev. B}\ }\textbf {\bibinfo {volume} {108}},\ \bibinfo
  {pages} {205144} (\bibinfo {year} {2023})}\BibitemShut {NoStop}%
\bibitem [{\citenamefont {Carleo}\ and\ \citenamefont
  {Troyer}(2017)}]{Carleo2017Solving}%
  \BibitemOpen
  \bibfield  {author} {\bibinfo {author} {\bibfnamefont {G.}~\bibnamefont
  {Carleo}}\ and\ \bibinfo {author} {\bibfnamefont {M.}~\bibnamefont
  {Troyer}},\ }\bibfield  {title} {\bibinfo {title} {Solving the quantum
  many-body problem with artificial neural networks},\ }\href
  {https://doi.org/10.1126/science.aag2302} {\bibfield  {journal} {\bibinfo
  {journal} {Science}\ }\textbf {\bibinfo {volume} {355}},\ \bibinfo {pages}
  {602} (\bibinfo {year} {2017})}\BibitemShut {NoStop}%
\bibitem [{\citenamefont {Pfau}\ \emph {et~al.}(2020)\citenamefont {Pfau},
  \citenamefont {Spencer}, \citenamefont {Matthews},\ and\ \citenamefont
  {Foulkes}}]{Pfau2020Abinitio}%
  \BibitemOpen
  \bibfield  {author} {\bibinfo {author} {\bibfnamefont {D.}~\bibnamefont
  {Pfau}}, \bibinfo {author} {\bibfnamefont {J.~S.}\ \bibnamefont {Spencer}},
  \bibinfo {author} {\bibfnamefont {A.~G. D.~G.}\ \bibnamefont {Matthews}},\
  and\ \bibinfo {author} {\bibfnamefont {W.~M.~C.}\ \bibnamefont {Foulkes}},\
  }\bibfield  {title} {\bibinfo {title} {Ab initio solution of the
  many-electron {S}chr\"odinger equation with deep neural networks},\ }\href
  {https://doi.org/10.1103/PhysRevResearch.2.033429} {\bibfield  {journal}
  {\bibinfo  {journal} {Phys. Rev. Res.}\ }\textbf {\bibinfo {volume} {2}},\
  \bibinfo {pages} {033429} (\bibinfo {year} {2020})}\BibitemShut {NoStop}%
\bibitem [{\citenamefont {Pescia}\ \emph {et~al.}(2024)\citenamefont {Pescia},
  \citenamefont {Nys}, \citenamefont {Kim}, \citenamefont {Lovato},\ and\
  \citenamefont {Carleo}}]{Pescia2024Message}%
  \BibitemOpen
  \bibfield  {author} {\bibinfo {author} {\bibfnamefont {G.}~\bibnamefont
  {Pescia}}, \bibinfo {author} {\bibfnamefont {J.}~\bibnamefont {Nys}},
  \bibinfo {author} {\bibfnamefont {J.}~\bibnamefont {Kim}}, \bibinfo {author}
  {\bibfnamefont {A.}~\bibnamefont {Lovato}},\ and\ \bibinfo {author}
  {\bibfnamefont {G.}~\bibnamefont {Carleo}},\ }\bibfield  {title} {\bibinfo
  {title} {Message-passing neural quantum states for the homogeneous electron
  gas},\ }\href {https://doi.org/10.1103/PhysRevB.110.035108} {\bibfield
  {journal} {\bibinfo  {journal} {Phys. Rev. B}\ }\textbf {\bibinfo {volume}
  {110}},\ \bibinfo {pages} {035108} (\bibinfo {year} {2024})}\BibitemShut
  {NoStop}%
\bibitem [{\citenamefont {Smith}\ \emph {et~al.}(2024)\citenamefont {Smith},
  \citenamefont {Chen}, \citenamefont {Levy}, \citenamefont {Yang},
  \citenamefont {Morales},\ and\ \citenamefont {Zhang}}]{Smith2024WC}%
  \BibitemOpen
  \bibfield  {author} {\bibinfo {author} {\bibfnamefont {C.}~\bibnamefont
  {Smith}}, \bibinfo {author} {\bibfnamefont {Y.}~\bibnamefont {Chen}},
  \bibinfo {author} {\bibfnamefont {R.}~\bibnamefont {Levy}}, \bibinfo {author}
  {\bibfnamefont {Y.}~\bibnamefont {Yang}}, \bibinfo {author} {\bibfnamefont
  {M.~A.}\ \bibnamefont {Morales}},\ and\ \bibinfo {author} {\bibfnamefont
  {S.}~\bibnamefont {Zhang}},\ }\bibfield  {title} {\bibinfo {title} {Unified
  variational approach description of ground-state phases of the
  two-dimensional electron gas},\ }\href
  {https://doi.org/10.1103/PhysRevLett.133.266504} {\bibfield  {journal}
  {\bibinfo  {journal} {Phys. Rev. Lett.}\ }\textbf {\bibinfo {volume} {133}},\
  \bibinfo {pages} {266504} (\bibinfo {year} {2024})}\BibitemShut {NoStop}%
\bibitem [{\citenamefont {Viteritti}\ \emph {et~al.}(2023)\citenamefont
  {Viteritti}, \citenamefont {Rende},\ and\ \citenamefont
  {Becca}}]{Viteritti2023Transformer}%
  \BibitemOpen
  \bibfield  {author} {\bibinfo {author} {\bibfnamefont {L.~L.}\ \bibnamefont
  {Viteritti}}, \bibinfo {author} {\bibfnamefont {R.}~\bibnamefont {Rende}},\
  and\ \bibinfo {author} {\bibfnamefont {F.}~\bibnamefont {Becca}},\ }\bibfield
   {title} {\bibinfo {title} {Transformer variational wave functions for
  frustrated quantum spin systems},\ }\href
  {https://doi.org/10.1103/PhysRevLett.130.236401} {\bibfield  {journal}
  {\bibinfo  {journal} {Phys. Rev. Lett.}\ }\textbf {\bibinfo {volume} {130}},\
  \bibinfo {pages} {236401} (\bibinfo {year} {2023})}\BibitemShut {NoStop}%
\bibitem [{\citenamefont {Lange}\ \emph {et~al.}(2026)\citenamefont {Lange},
  \citenamefont {Chen}, \citenamefont {Georges}, \citenamefont {Grusdt},
  \citenamefont {Bohrdt},\ and\ \citenamefont {Roth}}]{lange2026simulating}%
  \BibitemOpen
  \bibfield  {author} {\bibinfo {author} {\bibfnamefont {H.}~\bibnamefont
  {Lange}}, \bibinfo {author} {\bibfnamefont {A.}~\bibnamefont {Chen}},
  \bibinfo {author} {\bibfnamefont {A.}~\bibnamefont {Georges}}, \bibinfo
  {author} {\bibfnamefont {F.}~\bibnamefont {Grusdt}}, \bibinfo {author}
  {\bibfnamefont {A.}~\bibnamefont {Bohrdt}},\ and\ \bibinfo {author}
  {\bibfnamefont {C.}~\bibnamefont {Roth}},\ }\href
  {https://arxiv.org/abs/2602.10091} {\bibinfo {title} {Simulating
  superconductivity in mixed-dimensional
  $t_\parallel$-${J}_\parallel$-${J}_\perp$ bilayers with neural quantum
  states}} (\bibinfo {year} {2026}),\ \Eprint
  {https://arxiv.org/abs/2602.10091} {arXiv:2602.10091 [cond-mat.str-el]}
  \BibitemShut {NoStop}%
\bibitem [{\citenamefont {Roth}\ \emph {et~al.}(2025)\citenamefont {Roth},
  \citenamefont {Chen}, \citenamefont {Sengupta},\ and\ \citenamefont
  {Georges}}]{roth2025superconductivity}%
  \BibitemOpen
  \bibfield  {author} {\bibinfo {author} {\bibfnamefont {C.}~\bibnamefont
  {Roth}}, \bibinfo {author} {\bibfnamefont {A.}~\bibnamefont {Chen}}, \bibinfo
  {author} {\bibfnamefont {A.}~\bibnamefont {Sengupta}},\ and\ \bibinfo
  {author} {\bibfnamefont {A.}~\bibnamefont {Georges}},\ }\href
  {https://arxiv.org/abs/2511.07566} {\bibinfo {title} {Superconductivity in
  the two-dimensional {H}ubbard model revealed by neural quantum states}}
  (\bibinfo {year} {2025}),\ \Eprint {https://arxiv.org/abs/2511.07566}
  {arXiv:2511.07566 [cond-mat.supr-con]} \BibitemShut {NoStop}%
\bibitem [{\citenamefont {Viteritti}\ \emph
  {et~al.}(2026{\natexlab{a}})\citenamefont {Viteritti}, \citenamefont {Rende},
  \citenamefont {Sachdev},\ and\ \citenamefont
  {Carleo}}]{viteritti2026approaching}%
  \BibitemOpen
  \bibfield  {author} {\bibinfo {author} {\bibfnamefont {L.~L.}\ \bibnamefont
  {Viteritti}}, \bibinfo {author} {\bibfnamefont {R.}~\bibnamefont {Rende}},
  \bibinfo {author} {\bibfnamefont {S.}~\bibnamefont {Sachdev}},\ and\ \bibinfo
  {author} {\bibfnamefont {G.}~\bibnamefont {Carleo}},\ }\href
  {https://arxiv.org/abs/2602.02665} {\bibinfo {title} {Approaching the
  thermodynamic limit with neural-network quantum states}} (\bibinfo {year}
  {2026}{\natexlab{a}}),\ \Eprint {https://arxiv.org/abs/2602.02665}
  {arXiv:2602.02665 [cond-mat.str-el]} \BibitemShut {NoStop}%
\bibitem [{\citenamefont {Viteritti}\ \emph
  {et~al.}(2026{\natexlab{b}})\citenamefont {Viteritti}, \citenamefont {Rende},
  \citenamefont {Roth}, \citenamefont {Sengupta}, \citenamefont {Carleo},\ and\
  \citenamefont {Georges}}]{viteritti2026variational}%
  \BibitemOpen
  \bibfield  {author} {\bibinfo {author} {\bibfnamefont {L.~L.}\ \bibnamefont
  {Viteritti}}, \bibinfo {author} {\bibfnamefont {R.}~\bibnamefont {Rende}},
  \bibinfo {author} {\bibfnamefont {C.}~\bibnamefont {Roth}}, \bibinfo {author}
  {\bibfnamefont {A.}~\bibnamefont {Sengupta}}, \bibinfo {author}
  {\bibfnamefont {G.}~\bibnamefont {Carleo}},\ and\ \bibinfo {author}
  {\bibfnamefont {A.}~\bibnamefont {Georges}},\ }\href
  {https://arxiv.org/abs/2604.21978} {\bibinfo {title} {Beyond variational
  bias: Resolving intertwined orders in the {H}ubbard model}} (\bibinfo {year}
  {2026}{\natexlab{b}}),\ \Eprint {https://arxiv.org/abs/2604.21978}
  {arXiv:2604.21978 [cond-mat.str-el]} \BibitemShut {NoStop}%
\bibitem [{\citenamefont {Roth}\ \emph {et~al.}(2026)\citenamefont {Roth},
  \citenamefont {Millis},\ and\ \citenamefont {Soejima}}]{roth2026largescale}%
  \BibitemOpen
  \bibfield  {author} {\bibinfo {author} {\bibfnamefont {C.}~\bibnamefont
  {Roth}}, \bibinfo {author} {\bibfnamefont {A.}~\bibnamefont {Millis}},\ and\
  \bibinfo {author} {\bibfnamefont {T.}~\bibnamefont {Soejima}},\ }\href
  {https://arxiv.org/abs/2608.02753} {\bibinfo {title} {Large scale neural
  quantum states reveal the interplay between superconductivity and quantum
  criticality in the {H}ofstadter-{H}ubbard model}} (\bibinfo {year} {2026}),\
  \Eprint {https://arxiv.org/abs/2608.02753} {arXiv:2608.02753
  [cond-mat.str-el]} \BibitemShut {NoStop}%
\bibitem [{\citenamefont {Qian}\ \emph {et~al.}(2025)\citenamefont {Qian},
  \citenamefont {Zhao}, \citenamefont {Zhang}, \citenamefont {Xiang},
  \citenamefont {Li},\ and\ \citenamefont {Chen}}]{Qian2025Describing}%
  \BibitemOpen
  \bibfield  {author} {\bibinfo {author} {\bibfnamefont {Y.}~\bibnamefont
  {Qian}}, \bibinfo {author} {\bibfnamefont {T.}~\bibnamefont {Zhao}}, \bibinfo
  {author} {\bibfnamefont {J.}~\bibnamefont {Zhang}}, \bibinfo {author}
  {\bibfnamefont {T.}~\bibnamefont {Xiang}}, \bibinfo {author} {\bibfnamefont
  {X.}~\bibnamefont {Li}},\ and\ \bibinfo {author} {\bibfnamefont
  {J.}~\bibnamefont {Chen}},\ }\bibfield  {title} {\bibinfo {title} {Describing
  {L}andau level mixing in fractional quantum {H}all states with deep
  learning},\ }\href {https://doi.org/10.1103/PhysRevLett.134.176503}
  {\bibfield  {journal} {\bibinfo  {journal} {Phys. Rev. Lett.}\ }\textbf
  {\bibinfo {volume} {134}},\ \bibinfo {pages} {176503} (\bibinfo {year}
  {2025})}\BibitemShut {NoStop}%
\bibitem [{\citenamefont {Teng}\ \emph {et~al.}(2025)\citenamefont {Teng},
  \citenamefont {Dai},\ and\ \citenamefont {Fu}}]{Teng2025Solving}%
  \BibitemOpen
  \bibfield  {author} {\bibinfo {author} {\bibfnamefont {Y.}~\bibnamefont
  {Teng}}, \bibinfo {author} {\bibfnamefont {D.~D.}\ \bibnamefont {Dai}},\ and\
  \bibinfo {author} {\bibfnamefont {L.}~\bibnamefont {Fu}},\ }\bibfield
  {title} {\bibinfo {title} {Solving the fractional quantum {H}all problem with
  self-attention neural network},\ }\href
  {https://doi.org/10.1103/PhysRevB.111.205117} {\bibfield  {journal} {\bibinfo
   {journal} {Phys. Rev. B}\ }\textbf {\bibinfo {volume} {111}},\ \bibinfo
  {pages} {205117} (\bibinfo {year} {2025})}\BibitemShut {NoStop}%
\bibitem [{\citenamefont {Gattu}(2026)}]{Gattu2026Dressing}%
  \BibitemOpen
  \bibfield  {author} {\bibinfo {author} {\bibfnamefont {M.}~\bibnamefont
  {Gattu}},\ }\bibfield  {title} {\bibinfo {title} {Dressing composite fermions
  with artificial intelligence},\ }\href {https://doi.org/10.1103/zq76-147c}
  {\bibfield  {journal} {\bibinfo  {journal} {PRX Intelligence}\ }\textbf
  {\bibinfo {volume} {1}},\ \bibinfo {pages} {023009} (\bibinfo {year}
  {2026})}\BibitemShut {NoStop}%
\bibitem [{\citenamefont {Zhu}\ \emph {et~al.}(2026)\citenamefont {Zhu},
  \citenamefont {Huang}, \citenamefont {Hu}, \citenamefont {Xiao},\ and\
  \citenamefont {Cao}}]{zhu2026crystallization}%
  \BibitemOpen
  \bibfield  {author} {\bibinfo {author} {\bibfnamefont {J.}~\bibnamefont
  {Zhu}}, \bibinfo {author} {\bibfnamefont {Y.}~\bibnamefont {Huang}}, \bibinfo
  {author} {\bibfnamefont {X.}~\bibnamefont {Hu}}, \bibinfo {author}
  {\bibfnamefont {D.}~\bibnamefont {Xiao}},\ and\ \bibinfo {author}
  {\bibfnamefont {T.}~\bibnamefont {Cao}},\ }\href
  {https://arxiv.org/abs/2604.06316} {\bibinfo {title} {Crystallization in the
  fractional quantum {H}all regime with disorder-aware neural quantum states}}
  (\bibinfo {year} {2026}),\ \Eprint {https://arxiv.org/abs/2604.06316}
  {arXiv:2604.06316 [cond-mat.str-el]} \BibitemShut {NoStop}%
\bibitem [{\citenamefont {Graham}\ \emph {et~al.}(2026)\citenamefont {Graham},
  \citenamefont {Abouelkomsan},\ and\ \citenamefont
  {Fu}}]{graham2026nonabelian}%
  \BibitemOpen
  \bibfield  {author} {\bibinfo {author} {\bibfnamefont {P.-A.}\ \bibnamefont
  {Graham}}, \bibinfo {author} {\bibfnamefont {A.}~\bibnamefont
  {Abouelkomsan}},\ and\ \bibinfo {author} {\bibfnamefont {L.}~\bibnamefont
  {Fu}},\ }\href {https://arxiv.org/abs/2610.18967} {\bibinfo {title}
  {Non-{A}belian quantum {H}all states at strong {L}andau level mixing: finding
  anyons with first-principles {AI}}} (\bibinfo {year} {2026}),\ \Eprint
  {https://arxiv.org/abs/2610.18967} {arXiv:2610.18967 [cond-mat.mes-hall]}
  \BibitemShut {NoStop}%
\bibitem [{\citenamefont {Perez~Fadon}\ \emph {et~al.}(2026)\citenamefont
  {Perez~Fadon}, \citenamefont {Pfau}, \citenamefont {Spencer}, \citenamefont
  {Lou}, \citenamefont {Neupert},\ and\ \citenamefont
  {Foulkes}}]{Perez2026Extracting}%
  \BibitemOpen
  \bibfield  {author} {\bibinfo {author} {\bibfnamefont {A.}~\bibnamefont
  {Perez~Fadon}}, \bibinfo {author} {\bibfnamefont {D.}~\bibnamefont {Pfau}},
  \bibinfo {author} {\bibfnamefont {J.~S.}\ \bibnamefont {Spencer}}, \bibinfo
  {author} {\bibfnamefont {W.~T.}\ \bibnamefont {Lou}}, \bibinfo {author}
  {\bibfnamefont {T.}~\bibnamefont {Neupert}},\ and\ \bibinfo {author}
  {\bibfnamefont {W.~M.~C.}\ \bibnamefont {Foulkes}},\ }\bibfield  {title}
  {\bibinfo {title} {Extracting anyon statistics from neural network fractional
  quantum {H}all states},\ }\href {https://doi.org/10.1103/sb8g-qvn6}
  {\bibfield  {journal} {\bibinfo  {journal} {PRX Intelligence}\ }\textbf
  {\bibinfo {volume} {1}},\ \bibinfo {pages} {013018} (\bibinfo {year}
  {2026})}\BibitemShut {NoStop}%
\bibitem [{\citenamefont {Abouelkomsan}\ and\ \citenamefont
  {Fu}(2026)}]{Abouelkomsan2026First}%
  \BibitemOpen
  \bibfield  {author} {\bibinfo {author} {\bibfnamefont {A.}~\bibnamefont
  {Abouelkomsan}}\ and\ \bibinfo {author} {\bibfnamefont {L.}~\bibnamefont
  {Fu}},\ }\bibfield  {title} {\bibinfo {title} {First-principles {AI} finds
  crystallization of fractional quantum {H}all liquids},\ }\href
  {https://doi.org/10.1103/9qlr-jp6x} {\bibfield  {journal} {\bibinfo
  {journal} {PRX Intelligence}\ }\textbf {\bibinfo {volume} {1}},\ \bibinfo
  {pages} {013010} (\bibinfo {year} {2026})}\BibitemShut {NoStop}%
\bibitem [{\citenamefont {Abouelkomsan}\ \emph
  {et~al.}(2026{\natexlab{a}})\citenamefont {Abouelkomsan}, \citenamefont
  {Geier},\ and\ \citenamefont {Fu}}]{Abouelkomsan2026Topological}%
  \BibitemOpen
  \bibfield  {author} {\bibinfo {author} {\bibfnamefont {A.}~\bibnamefont
  {Abouelkomsan}}, \bibinfo {author} {\bibfnamefont {M.}~\bibnamefont
  {Geier}},\ and\ \bibinfo {author} {\bibfnamefont {L.}~\bibnamefont {Fu}},\
  }\bibfield  {title} {\bibinfo {title} {Topological order in neural
  wavefunctions},\ }\href {https://doi.org/10.1103/bzq1-123h} {\bibfield
  {journal} {\bibinfo  {journal} {Phys. Rev. B}\ }\textbf {\bibinfo {volume}
  {113}},\ \bibinfo {pages} {205119} (\bibinfo {year}
  {2026}{\natexlab{a}})}\BibitemShut {NoStop}%
\bibitem [{\citenamefont {Li}\ \emph {et~al.}(2025)\citenamefont {Li},
  \citenamefont {Chen}, \citenamefont {Li}, \citenamefont {Chen}, \citenamefont
  {Wu}, \citenamefont {Chen},\ and\ \citenamefont {Ren}}]{li2025deep}%
  \BibitemOpen
  \bibfield  {author} {\bibinfo {author} {\bibfnamefont {X.}~\bibnamefont
  {Li}}, \bibinfo {author} {\bibfnamefont {Y.}~\bibnamefont {Chen}}, \bibinfo
  {author} {\bibfnamefont {B.}~\bibnamefont {Li}}, \bibinfo {author}
  {\bibfnamefont {H.}~\bibnamefont {Chen}}, \bibinfo {author} {\bibfnamefont
  {F.}~\bibnamefont {Wu}}, \bibinfo {author} {\bibfnamefont {J.}~\bibnamefont
  {Chen}},\ and\ \bibinfo {author} {\bibfnamefont {W.}~\bibnamefont {Ren}},\
  }\href {https://arxiv.org/abs/2503.11756} {\bibinfo {title} {Deep learning
  sheds light on integer and fractional topological insulators}} (\bibinfo
  {year} {2025}),\ \Eprint {https://arxiv.org/abs/2503.11756} {arXiv:2503.11756
  [cond-mat.str-el]} \BibitemShut {NoStop}%
\bibitem [{\citenamefont {Luo}\ \emph {et~al.}(2025)\citenamefont {Luo},
  \citenamefont {Zaklama},\ and\ \citenamefont {Fu}}]{luo2025solving}%
  \BibitemOpen
  \bibfield  {author} {\bibinfo {author} {\bibfnamefont {D.}~\bibnamefont
  {Luo}}, \bibinfo {author} {\bibfnamefont {T.}~\bibnamefont {Zaklama}},\ and\
  \bibinfo {author} {\bibfnamefont {L.}~\bibnamefont {Fu}},\ }\href
  {https://arxiv.org/abs/2503.13585} {\bibinfo {title} {Solving fractional
  electron states in twisted {M}o{T}e$_2$ with deep neural network}} (\bibinfo
  {year} {2025}),\ \Eprint {https://arxiv.org/abs/2503.13585} {arXiv:2503.13585
  [cond-mat.str-el]} \BibitemShut {NoStop}%
\bibitem [{\citenamefont {Smith}\ \emph {et~al.}(2026)\citenamefont {Smith},
  \citenamefont {Yang}, \citenamefont {Wan}, \citenamefont {Chen},
  \citenamefont {Morales},\ and\ \citenamefont {Zhang}}]{smith2026neural}%
  \BibitemOpen
  \bibfield  {author} {\bibinfo {author} {\bibfnamefont {C.}~\bibnamefont
  {Smith}}, \bibinfo {author} {\bibfnamefont {Y.}~\bibnamefont {Yang}},
  \bibinfo {author} {\bibfnamefont {Z.-Q.}\ \bibnamefont {Wan}}, \bibinfo
  {author} {\bibfnamefont {Y.}~\bibnamefont {Chen}}, \bibinfo {author}
  {\bibfnamefont {M.~A.}\ \bibnamefont {Morales}},\ and\ \bibinfo {author}
  {\bibfnamefont {S.}~\bibnamefont {Zhang}},\ }\href
  {https://arxiv.org/abs/2602.16798} {\bibinfo {title} {Neural network
  discovery of paired {W}igner crystals in artificial graphene}} (\bibinfo
  {year} {2026}),\ \Eprint {https://arxiv.org/abs/2602.16798} {arXiv:2602.16798
  [quant-ph]} \BibitemShut {NoStop}%
\bibitem [{\citenamefont {Geier}\ \emph {et~al.}(2025)\citenamefont {Geier},
  \citenamefont {Nazaryan}, \citenamefont {Zaklama},\ and\ \citenamefont
  {Fu}}]{Geier2025Self}%
  \BibitemOpen
  \bibfield  {author} {\bibinfo {author} {\bibfnamefont {M.}~\bibnamefont
  {Geier}}, \bibinfo {author} {\bibfnamefont {K.}~\bibnamefont {Nazaryan}},
  \bibinfo {author} {\bibfnamefont {T.}~\bibnamefont {Zaklama}},\ and\ \bibinfo
  {author} {\bibfnamefont {L.}~\bibnamefont {Fu}},\ }\bibfield  {title}
  {\bibinfo {title} {Self-attention neural network for solving correlated
  electron problems in solids},\ }\href {https://doi.org/10.1103/qxc3-bkc7}
  {\bibfield  {journal} {\bibinfo  {journal} {Phys. Rev. B}\ }\textbf {\bibinfo
  {volume} {112}},\ \bibinfo {pages} {045119} (\bibinfo {year}
  {2025})}\BibitemShut {NoStop}%
\bibitem [{\citenamefont {Li}\ \emph {et~al.}(2024{\natexlab{a}})\citenamefont
  {Li}, \citenamefont {Qian}, \citenamefont {Ren}, \citenamefont {Xu},\ and\
  \citenamefont {Chen}}]{li2024emergent}%
  \BibitemOpen
  \bibfield  {author} {\bibinfo {author} {\bibfnamefont {X.}~\bibnamefont
  {Li}}, \bibinfo {author} {\bibfnamefont {Y.}~\bibnamefont {Qian}}, \bibinfo
  {author} {\bibfnamefont {W.}~\bibnamefont {Ren}}, \bibinfo {author}
  {\bibfnamefont {Y.}~\bibnamefont {Xu}},\ and\ \bibinfo {author}
  {\bibfnamefont {J.}~\bibnamefont {Chen}},\ }\href
  {https://arxiv.org/abs/2406.11134} {\bibinfo {title} {Emergent {W}igner
  phases in moir\'e superlattice from deep learning}} (\bibinfo {year}
  {2024}{\natexlab{a}}),\ \Eprint {https://arxiv.org/abs/2406.11134}
  {arXiv:2406.11134 [physics.comp-ph]} \BibitemShut {NoStop}%
\bibitem [{\citenamefont {Valenti}\ \emph {et~al.}(2025)\citenamefont
  {Valenti}, \citenamefont {Vituri}, \citenamefont {Yang}, \citenamefont
  {Parker}, \citenamefont {Soejima}, \citenamefont {Dong}, \citenamefont
  {Morales}, \citenamefont {Vishwanath}, \citenamefont {Berg},\ and\
  \citenamefont {Zhang}}]{valenti2025quantum}%
  \BibitemOpen
  \bibfield  {author} {\bibinfo {author} {\bibfnamefont {A.}~\bibnamefont
  {Valenti}}, \bibinfo {author} {\bibfnamefont {Y.}~\bibnamefont {Vituri}},
  \bibinfo {author} {\bibfnamefont {Y.}~\bibnamefont {Yang}}, \bibinfo {author}
  {\bibfnamefont {D.~E.}\ \bibnamefont {Parker}}, \bibinfo {author}
  {\bibfnamefont {T.}~\bibnamefont {Soejima}}, \bibinfo {author} {\bibfnamefont
  {J.}~\bibnamefont {Dong}}, \bibinfo {author} {\bibfnamefont {M.~A.}\
  \bibnamefont {Morales}}, \bibinfo {author} {\bibfnamefont {A.}~\bibnamefont
  {Vishwanath}}, \bibinfo {author} {\bibfnamefont {E.}~\bibnamefont {Berg}},\
  and\ \bibinfo {author} {\bibfnamefont {S.}~\bibnamefont {Zhang}},\ }\href
  {https://arxiv.org/abs/2512.07947} {\bibinfo {title} {Quantum geometry driven
  crystallization: A neural-network variational {M}onte {C}arlo study}}
  (\bibinfo {year} {2025}),\ \Eprint {https://arxiv.org/abs/2512.07947}
  {arXiv:2512.07947 [cond-mat.str-el]} \BibitemShut {NoStop}%
\bibitem [{\citenamefont {Abouelkomsan}\ \emph
  {et~al.}(2026{\natexlab{b}})\citenamefont {Abouelkomsan}, \citenamefont
  {Gaggioli}, \citenamefont {Guerci},\ and\ \citenamefont
  {Fu}}]{abouelkomsan2026rhombohedral}%
  \BibitemOpen
  \bibfield  {author} {\bibinfo {author} {\bibfnamefont {A.}~\bibnamefont
  {Abouelkomsan}}, \bibinfo {author} {\bibfnamefont {F.}~\bibnamefont
  {Gaggioli}}, \bibinfo {author} {\bibfnamefont {D.}~\bibnamefont {Guerci}},\
  and\ \bibinfo {author} {\bibfnamefont {L.}~\bibnamefont {Fu}},\ }\href
  {https://arxiv.org/abs/2608.00167} {\bibinfo {title} {Rhombohedral graphene:
  A tale of many crystals}} (\bibinfo {year} {2026}{\natexlab{b}}),\ \Eprint
  {https://arxiv.org/abs/2608.00167} {arXiv:2608.00167 [cond-mat.str-el]}
  \BibitemShut {NoStop}%
\bibitem [{\citenamefont {Aharonov}\ and\ \citenamefont
  {Casher}(1979)}]{aharonov1979ground}%
  \BibitemOpen
  \bibfield  {author} {\bibinfo {author} {\bibfnamefont {Y.}~\bibnamefont
  {Aharonov}}\ and\ \bibinfo {author} {\bibfnamefont {A.}~\bibnamefont
  {Casher}},\ }\bibfield  {title} {\bibinfo {title} {Ground state of a
  spin-{$\frac{1}{2}$} charged particle in a two-dimensional magnetic field},\
  }\href {https://doi.org/10.1103/PhysRevA.19.2461} {\bibfield  {journal}
  {\bibinfo  {journal} {Phys. Rev. A}\ }\textbf {\bibinfo {volume} {19}},\
  \bibinfo {pages} {2461} (\bibinfo {year} {1979})}\BibitemShut {NoStop}%
\bibitem [{\citenamefont {Morales-Dur\'an}\ \emph {et~al.}(2024)\citenamefont
  {Morales-Dur\'an}, \citenamefont {Wei}, \citenamefont {Shi},\ and\
  \citenamefont {MacDonald}}]{Duran2024Magic}%
  \BibitemOpen
  \bibfield  {author} {\bibinfo {author} {\bibfnamefont {N.}~\bibnamefont
  {Morales-Dur\'an}}, \bibinfo {author} {\bibfnamefont {N.}~\bibnamefont
  {Wei}}, \bibinfo {author} {\bibfnamefont {J.}~\bibnamefont {Shi}},\ and\
  \bibinfo {author} {\bibfnamefont {A.~H.}\ \bibnamefont {MacDonald}},\
  }\bibfield  {title} {\bibinfo {title} {Magic angles and fractional {C}hern
  insulators in twisted homobilayer transition metal dichalcogenides},\ }\href
  {https://doi.org/10.1103/PhysRevLett.132.096602} {\bibfield  {journal}
  {\bibinfo  {journal} {Phys. Rev. Lett.}\ }\textbf {\bibinfo {volume} {132}},\
  \bibinfo {pages} {096602} (\bibinfo {year} {2024})}\BibitemShut {NoStop}%
\bibitem [{\citenamefont {Shi}\ \emph {et~al.}(2024)\citenamefont {Shi},
  \citenamefont {Morales-Dur\'an}, \citenamefont {Khalaf},\ and\ \citenamefont
  {MacDonald}}]{Shi2024Adiabatic}%
  \BibitemOpen
  \bibfield  {author} {\bibinfo {author} {\bibfnamefont {J.}~\bibnamefont
  {Shi}}, \bibinfo {author} {\bibfnamefont {N.}~\bibnamefont
  {Morales-Dur\'an}}, \bibinfo {author} {\bibfnamefont {E.}~\bibnamefont
  {Khalaf}},\ and\ \bibinfo {author} {\bibfnamefont {A.~H.}\ \bibnamefont
  {MacDonald}},\ }\bibfield  {title} {\bibinfo {title} {Adiabatic approximation
  and {A}haronov-{C}asher bands in twisted homobilayer transition metal
  dichalcogenides},\ }\href {https://doi.org/10.1103/PhysRevB.110.035130}
  {\bibfield  {journal} {\bibinfo  {journal} {Phys. Rev. B}\ }\textbf {\bibinfo
  {volume} {110}},\ \bibinfo {pages} {035130} (\bibinfo {year}
  {2024})}\BibitemShut {NoStop}%
\bibitem [{\citenamefont {Tang}\ \emph {et~al.}(2011)\citenamefont {Tang},
  \citenamefont {Mei},\ and\ \citenamefont {Wen}}]{Wen_FCI}%
  \BibitemOpen
  \bibfield  {author} {\bibinfo {author} {\bibfnamefont {E.}~\bibnamefont
  {Tang}}, \bibinfo {author} {\bibfnamefont {J.-W.}\ \bibnamefont {Mei}},\ and\
  \bibinfo {author} {\bibfnamefont {X.-G.}\ \bibnamefont {Wen}},\ }\bibfield
  {title} {\bibinfo {title} {High-temperature fractional quantum {H}all
  states},\ }\href {https://doi.org/10.1103/PhysRevLett.106.236802} {\bibfield
  {journal} {\bibinfo  {journal} {Phys. Rev. Lett.}\ }\textbf {\bibinfo
  {volume} {106}},\ \bibinfo {pages} {236802} (\bibinfo {year}
  {2011})}\BibitemShut {NoStop}%
\bibitem [{\citenamefont {Neupert}\ \emph {et~al.}(2011)\citenamefont
  {Neupert}, \citenamefont {Santos}, \citenamefont {Chamon},\ and\
  \citenamefont {Mudry}}]{Neupert_FCI}%
  \BibitemOpen
  \bibfield  {author} {\bibinfo {author} {\bibfnamefont {T.}~\bibnamefont
  {Neupert}}, \bibinfo {author} {\bibfnamefont {L.}~\bibnamefont {Santos}},
  \bibinfo {author} {\bibfnamefont {C.}~\bibnamefont {Chamon}},\ and\ \bibinfo
  {author} {\bibfnamefont {C.}~\bibnamefont {Mudry}},\ }\bibfield  {title}
  {\bibinfo {title} {Fractional quantum {H}all states at zero magnetic field},\
  }\href {https://doi.org/10.1103/PhysRevLett.106.236804} {\bibfield  {journal}
  {\bibinfo  {journal} {Phys. Rev. Lett.}\ }\textbf {\bibinfo {volume} {106}},\
  \bibinfo {pages} {236804} (\bibinfo {year} {2011})}\BibitemShut {NoStop}%
\bibitem [{\citenamefont {Sun}\ \emph {et~al.}(2011)\citenamefont {Sun},
  \citenamefont {Gu}, \citenamefont {Katsura},\ and\ \citenamefont
  {Das~Sarma}}]{DasSarma_Sun_FCI}%
  \BibitemOpen
  \bibfield  {author} {\bibinfo {author} {\bibfnamefont {K.}~\bibnamefont
  {Sun}}, \bibinfo {author} {\bibfnamefont {Z.}~\bibnamefont {Gu}}, \bibinfo
  {author} {\bibfnamefont {H.}~\bibnamefont {Katsura}},\ and\ \bibinfo {author}
  {\bibfnamefont {S.}~\bibnamefont {Das~Sarma}},\ }\bibfield  {title} {\bibinfo
  {title} {Nearly flatbands with nontrivial topology},\ }\href
  {https://doi.org/10.1103/PhysRevLett.106.236803} {\bibfield  {journal}
  {\bibinfo  {journal} {Phys. Rev. Lett.}\ }\textbf {\bibinfo {volume} {106}},\
  \bibinfo {pages} {236803} (\bibinfo {year} {2011})}\BibitemShut {NoStop}%
\bibitem [{\citenamefont {Regnault}\ and\ \citenamefont
  {Bernevig}(2011)}]{Bernevig_Regnault_FCI}%
  \BibitemOpen
  \bibfield  {author} {\bibinfo {author} {\bibfnamefont {N.}~\bibnamefont
  {Regnault}}\ and\ \bibinfo {author} {\bibfnamefont {B.~A.}\ \bibnamefont
  {Bernevig}},\ }\bibfield  {title} {\bibinfo {title} {Fractional {C}hern
  insulator},\ }\href {https://doi.org/10.1103/PhysRevX.1.021014} {\bibfield
  {journal} {\bibinfo  {journal} {Phys. Rev. X}\ }\textbf {\bibinfo {volume}
  {1}},\ \bibinfo {pages} {021014} (\bibinfo {year} {2011})}\BibitemShut
  {NoStop}%
\bibitem [{\citenamefont {Yoshioka}\ and\ \citenamefont
  {Fukuyama}(1979)}]{Yoshioka_Fukuyama}%
  \BibitemOpen
  \bibfield  {author} {\bibinfo {author} {\bibfnamefont {D.}~\bibnamefont
  {Yoshioka}}\ and\ \bibinfo {author} {\bibfnamefont {H.}~\bibnamefont
  {Fukuyama}},\ }\bibfield  {title} {\bibinfo {title} {Charge density wave
  state of two-dimensional electrons in strong magnetic fields},\ }\href
  {https://doi.org/10.1143/JPSJ.47.394} {\bibfield  {journal} {\bibinfo
  {journal} {Journal of the Physical Society of Japan}\ }\textbf {\bibinfo
  {volume} {47}},\ \bibinfo {pages} {394} (\bibinfo {year} {1979})}\BibitemShut
  {NoStop}%
\bibitem [{\citenamefont {MacDonald}\ and\ \citenamefont
  {Murray}(1985)}]{MacDonald1985Broken}%
  \BibitemOpen
  \bibfield  {author} {\bibinfo {author} {\bibfnamefont {A.~H.}\ \bibnamefont
  {MacDonald}}\ and\ \bibinfo {author} {\bibfnamefont {D.~B.}\ \bibnamefont
  {Murray}},\ }\bibfield  {title} {\bibinfo {title} {Broken symmetry states for
  two-dimensional electrons in a strong magnetic field},\ }\href
  {https://doi.org/10.1103/PhysRevB.32.2291} {\bibfield  {journal} {\bibinfo
  {journal} {Phys. Rev. B}\ }\textbf {\bibinfo {volume} {32}},\ \bibinfo
  {pages} {2291} (\bibinfo {year} {1985})}\BibitemShut {NoStop}%
\bibitem [{\citenamefont {Maki}\ and\ \citenamefont
  {Zotos}(1983)}]{maki1983static}%
  \BibitemOpen
  \bibfield  {author} {\bibinfo {author} {\bibfnamefont {K.}~\bibnamefont
  {Maki}}\ and\ \bibinfo {author} {\bibfnamefont {X.}~\bibnamefont {Zotos}},\
  }\bibfield  {title} {\bibinfo {title} {Static and dynamic properties of a
  two-dimensional {W}igner crystal in a strong magnetic field},\ }\href@noop {}
  {\bibfield  {journal} {\bibinfo  {journal} {Physical Review B}\ }\textbf
  {\bibinfo {volume} {28}},\ \bibinfo {pages} {4349} (\bibinfo {year}
  {1983})}\BibitemShut {NoStop}%
\bibitem [{\citenamefont {Sommer}\ and\ \citenamefont
  {Cooper}(2026)}]{sommer2026ideal}%
  \BibitemOpen
  \bibfield  {author} {\bibinfo {author} {\bibfnamefont {O.~E.}\ \bibnamefont
  {Sommer}}\ and\ \bibinfo {author} {\bibfnamefont {N.~R.}\ \bibnamefont
  {Cooper}},\ }\href {https://arxiv.org/abs/2509.01481} {\bibinfo {title}
  {Ideal optical flux lattices}} (\bibinfo {year} {2026}),\ \Eprint
  {https://arxiv.org/abs/2509.01481} {arXiv:2509.01481 [cond-mat.quant-gas]}
  \BibitemShut {NoStop}%
\bibitem [{\citenamefont {Wu}\ \emph {et~al.}(2019)\citenamefont {Wu},
  \citenamefont {Lovorn}, \citenamefont {Tutuc}, \citenamefont {Martin},\ and\
  \citenamefont {MacDonald}}]{Wu2019Topological}%
  \BibitemOpen
  \bibfield  {author} {\bibinfo {author} {\bibfnamefont {F.}~\bibnamefont
  {Wu}}, \bibinfo {author} {\bibfnamefont {T.}~\bibnamefont {Lovorn}}, \bibinfo
  {author} {\bibfnamefont {E.}~\bibnamefont {Tutuc}}, \bibinfo {author}
  {\bibfnamefont {I.}~\bibnamefont {Martin}},\ and\ \bibinfo {author}
  {\bibfnamefont {A.~H.}\ \bibnamefont {MacDonald}},\ }\bibfield  {title}
  {\bibinfo {title} {Topological insulators in twisted transition metal
  dichalcogenide homobilayers},\ }\href
  {https://doi.org/10.1103/PhysRevLett.122.086402} {\bibfield  {journal}
  {\bibinfo  {journal} {Phys. Rev. Lett.}\ }\textbf {\bibinfo {volume} {122}},\
  \bibinfo {pages} {086402} (\bibinfo {year} {2019})}\BibitemShut {NoStop}%
\bibitem [{\citenamefont {Shi}\ \emph {et~al.}(2025)\citenamefont {Shi},
  \citenamefont {Cano},\ and\ \citenamefont {Morales-Durán}}]{shi2025effects}%
  \BibitemOpen
  \bibfield  {author} {\bibinfo {author} {\bibfnamefont {J.}~\bibnamefont
  {Shi}}, \bibinfo {author} {\bibfnamefont {J.}~\bibnamefont {Cano}},\ and\
  \bibinfo {author} {\bibfnamefont {N.}~\bibnamefont {Morales-Durán}},\ }\href
  {https://arxiv.org/abs/2503.15900} {\bibinfo {title} {Effects of {B}erry
  curvature on ideal fractional {C}hern insulator many-body gaps}} (\bibinfo
  {year} {2025}),\ \Eprint {https://arxiv.org/abs/2503.15900} {arXiv:2503.15900
  [cond-mat.str-el]} \BibitemShut {NoStop}%
\bibitem [{\citenamefont {Iyer}\ \emph {et~al.}(2026)\citenamefont {Iyer},
  \citenamefont {Feuerpfeil}, \citenamefont {Crépel}, \citenamefont
  {Regnault},\ and\ \citenamefont {Mora}}]{iyer2026dispersion}%
  \BibitemOpen
  \bibfield  {author} {\bibinfo {author} {\bibfnamefont {K.}~\bibnamefont
  {Iyer}}, \bibinfo {author} {\bibfnamefont {A.}~\bibnamefont {Feuerpfeil}},
  \bibinfo {author} {\bibfnamefont {V.}~\bibnamefont {Crépel}}, \bibinfo
  {author} {\bibfnamefont {N.}~\bibnamefont {Regnault}},\ and\ \bibinfo
  {author} {\bibfnamefont {C.}~\bibnamefont {Mora}},\ }\href
  {https://arxiv.org/abs/2604.24859} {\bibinfo {title} {Dispersion of anyon
  {B}loch bands}} (\bibinfo {year} {2026}),\ \Eprint
  {https://arxiv.org/abs/2604.24859} {arXiv:2604.24859 [cond-mat.mes-hall]}
  \BibitemShut {NoStop}%
\bibitem [{\citenamefont {Morales-Durán}\ \emph
  {et~al.}(2026{\natexlab{a}})\citenamefont {Morales-Durán}, \citenamefont
  {Shi}, \citenamefont {Voinea}, \citenamefont {Potasz},\ and\ \citenamefont
  {Cano}}]{moralesduran2026bandmixing}%
  \BibitemOpen
  \bibfield  {author} {\bibinfo {author} {\bibfnamefont {N.}~\bibnamefont
  {Morales-Durán}}, \bibinfo {author} {\bibfnamefont {J.}~\bibnamefont {Shi}},
  \bibinfo {author} {\bibfnamefont {C.}~\bibnamefont {Voinea}}, \bibinfo
  {author} {\bibfnamefont {P.}~\bibnamefont {Potasz}},\ and\ \bibinfo {author}
  {\bibfnamefont {J.}~\bibnamefont {Cano}},\ }\href
  {https://arxiv.org/abs/2604.16847} {\bibinfo {title} {Band mixing and
  particle-hole asymmetry in moir\'e fractional {C}hern insulators}} (\bibinfo
  {year} {2026}{\natexlab{a}}),\ \Eprint {https://arxiv.org/abs/2604.16847}
  {arXiv:2604.16847 [cond-mat.str-el]} \BibitemShut {NoStop}%
\bibitem [{\citenamefont {Morales-Durán}\ \emph
  {et~al.}(2026{\natexlab{b}})\citenamefont {Morales-Durán}, \citenamefont
  {Wolf}, \citenamefont {Shi}, \citenamefont {Soejima}, \citenamefont
  {Millis},\ and\ \citenamefont {Cano}}]{Duran2026nonuniform}%
  \BibitemOpen
  \bibfield  {author} {\bibinfo {author} {\bibfnamefont {N.}~\bibnamefont
  {Morales-Durán}}, \bibinfo {author} {\bibfnamefont {T.~M.~R.}\ \bibnamefont
  {Wolf}}, \bibinfo {author} {\bibfnamefont {J.}~\bibnamefont {Shi}}, \bibinfo
  {author} {\bibfnamefont {T.}~\bibnamefont {Soejima}}, \bibinfo {author}
  {\bibfnamefont {A.~J.}\ \bibnamefont {Millis}},\ and\ \bibinfo {author}
  {\bibfnamefont {J.}~\bibnamefont {Cano}},\ }\href
  {https://arxiv.org/abs/2609.04149} {\bibinfo {title} {Non-uniform quantum
  geometry stabilizes generalized {W}igner crystals}} (\bibinfo {year}
  {2026}{\natexlab{b}}),\ \Eprint {https://arxiv.org/abs/2609.04149}
  {arXiv:2609.04149 [cond-mat.str-el]} \BibitemShut {NoStop}%
\bibitem [{\citenamefont {Guerci}\ \emph {et~al.}(2025)\citenamefont {Guerci},
  \citenamefont {Abouelkomsan},\ and\ \citenamefont {Fu}}]{Guerci_AC_SC}%
  \BibitemOpen
  \bibfield  {author} {\bibinfo {author} {\bibfnamefont {D.}~\bibnamefont
  {Guerci}}, \bibinfo {author} {\bibfnamefont {A.}~\bibnamefont
  {Abouelkomsan}},\ and\ \bibinfo {author} {\bibfnamefont {L.}~\bibnamefont
  {Fu}},\ }\bibfield  {title} {\bibinfo {title} {From fractionalization to
  chiral topological superconductivity in a flat {C}hern band},\ }\href
  {https://doi.org/10.1103/zm39-dstj} {\bibfield  {journal} {\bibinfo
  {journal} {Phys. Rev. Lett.}\ }\textbf {\bibinfo {volume} {135}},\ \bibinfo
  {pages} {186601} (\bibinfo {year} {2025})}\BibitemShut {NoStop}%
\bibitem [{\citenamefont {Jain}(1989)}]{Jain1989Composite}%
  \BibitemOpen
  \bibfield  {author} {\bibinfo {author} {\bibfnamefont {J.~K.}\ \bibnamefont
  {Jain}},\ }\bibfield  {title} {\bibinfo {title} {Composite-fermion approach
  for the fractional quantum {H}all effect},\ }\href
  {https://doi.org/10.1103/PhysRevLett.63.199} {\bibfield  {journal} {\bibinfo
  {journal} {Phys. Rev. Lett.}\ }\textbf {\bibinfo {volume} {63}},\ \bibinfo
  {pages} {199} (\bibinfo {year} {1989})}\BibitemShut {NoStop}%
\bibitem [{\citenamefont {Pu}\ \emph {et~al.}(2017)\citenamefont {Pu},
  \citenamefont {Wu},\ and\ \citenamefont {Jain}}]{Pu2017Composite}%
  \BibitemOpen
  \bibfield  {author} {\bibinfo {author} {\bibfnamefont {S.}~\bibnamefont
  {Pu}}, \bibinfo {author} {\bibfnamefont {Y.-H.}\ \bibnamefont {Wu}},\ and\
  \bibinfo {author} {\bibfnamefont {J.~K.}\ \bibnamefont {Jain}},\ }\bibfield
  {title} {\bibinfo {title} {Composite fermions on a torus},\ }\href
  {https://doi.org/10.1103/PhysRevB.96.195302} {\bibfield  {journal} {\bibinfo
  {journal} {Phys. Rev. B}\ }\textbf {\bibinfo {volume} {96}},\ \bibinfo
  {pages} {195302} (\bibinfo {year} {2017})}\BibitemShut {NoStop}%
\bibitem [{\citenamefont {Zhao}\ \emph {et~al.}(2018)\citenamefont {Zhao},
  \citenamefont {Zhang},\ and\ \citenamefont {Jain}}]{Zhao2018Crystallization}%
  \BibitemOpen
  \bibfield  {author} {\bibinfo {author} {\bibfnamefont {J.}~\bibnamefont
  {Zhao}}, \bibinfo {author} {\bibfnamefont {Y.}~\bibnamefont {Zhang}},\ and\
  \bibinfo {author} {\bibfnamefont {J.~K.}\ \bibnamefont {Jain}},\ }\bibfield
  {title} {\bibinfo {title} {Crystallization in the fractional quantum {H}all
  regime induced by {L}andau-level mixing},\ }\href
  {https://doi.org/10.1103/PhysRevLett.121.116802} {\bibfield  {journal}
  {\bibinfo  {journal} {Phys. Rev. Lett.}\ }\textbf {\bibinfo {volume} {121}},\
  \bibinfo {pages} {116802} (\bibinfo {year} {2018})}\BibitemShut {NoStop}%
\bibitem [{\citenamefont {Laughlin}(1983)}]{Laughlin1983Anomalous}%
  \BibitemOpen
  \bibfield  {author} {\bibinfo {author} {\bibfnamefont {R.~B.}\ \bibnamefont
  {Laughlin}},\ }\bibfield  {title} {\bibinfo {title} {Anomalous quantum {H}all
  effect: An incompressible quantum fluid with fractionally charged
  excitations},\ }\href {https://doi.org/10.1103/PhysRevLett.50.1395}
  {\bibfield  {journal} {\bibinfo  {journal} {Phys. Rev. Lett.}\ }\textbf
  {\bibinfo {volume} {50}},\ \bibinfo {pages} {1395} (\bibinfo {year}
  {1983})}\BibitemShut {NoStop}%
\bibitem [{\citenamefont {Platzman}\ and\ \citenamefont
  {Price}(1993)}]{Platzman1993QuantumFreezing}%
  \BibitemOpen
  \bibfield  {author} {\bibinfo {author} {\bibfnamefont {P.~M.}\ \bibnamefont
  {Platzman}}\ and\ \bibinfo {author} {\bibfnamefont {R.}~\bibnamefont
  {Price}},\ }\bibfield  {title} {\bibinfo {title} {Quantum freezing of the
  fractional quantum {H}all liquid},\ }\href
  {https://doi.org/10.1103/PhysRevLett.70.3487} {\bibfield  {journal} {\bibinfo
   {journal} {Phys. Rev. Lett.}\ }\textbf {\bibinfo {volume} {70}},\ \bibinfo
  {pages} {3487} (\bibinfo {year} {1993})}\BibitemShut {NoStop}%
\bibitem [{\citenamefont {Yi}\ and\ \citenamefont
  {Fertig}(1998)}]{Yi1998Laughlin}%
  \BibitemOpen
  \bibfield  {author} {\bibinfo {author} {\bibfnamefont {H.}~\bibnamefont
  {Yi}}\ and\ \bibinfo {author} {\bibfnamefont {H.~A.}\ \bibnamefont
  {Fertig}},\ }\bibfield  {title} {\bibinfo {title}
  {{L}aughlin-{J}astrow-correlated {W}igner crystal in a strong magnetic
  field},\ }\href {https://doi.org/10.1103/PhysRevB.58.4019} {\bibfield
  {journal} {\bibinfo  {journal} {Phys. Rev. B}\ }\textbf {\bibinfo {volume}
  {58}},\ \bibinfo {pages} {4019} (\bibinfo {year} {1998})}\BibitemShut
  {NoStop}%
\bibitem [{\citenamefont {Mishra}\ \emph {et~al.}(2026)\citenamefont {Mishra},
  \citenamefont {Wolf},\ and\ \citenamefont {MacDonald}}]{mishra2026charge}%
  \BibitemOpen
  \bibfield  {author} {\bibinfo {author} {\bibfnamefont {S.}~\bibnamefont
  {Mishra}}, \bibinfo {author} {\bibfnamefont {T.~M.~R.}\ \bibnamefont
  {Wolf}},\ and\ \bibinfo {author} {\bibfnamefont {A.~H.}\ \bibnamefont
  {MacDonald}},\ }\href {https://arxiv.org/abs/2603.05714} {\bibinfo {title}
  {Charge-ordered states in twisted {M}o{T}e$_2$}} (\bibinfo {year} {2026}),\
  \Eprint {https://arxiv.org/abs/2603.05714} {arXiv:2603.05714
  [cond-mat.str-el]} \BibitemShut {NoStop}%
\bibitem [{\citenamefont {L\"auchli}\ \emph {et~al.}(2013)\citenamefont
  {L\"auchli}, \citenamefont {Liu}, \citenamefont {Bergholtz},\ and\
  \citenamefont {Moessner}}]{Lauchli2013Hierarchy}%
  \BibitemOpen
  \bibfield  {author} {\bibinfo {author} {\bibfnamefont {A.~M.}\ \bibnamefont
  {L\"auchli}}, \bibinfo {author} {\bibfnamefont {Z.}~\bibnamefont {Liu}},
  \bibinfo {author} {\bibfnamefont {E.~J.}\ \bibnamefont {Bergholtz}},\ and\
  \bibinfo {author} {\bibfnamefont {R.}~\bibnamefont {Moessner}},\ }\bibfield
  {title} {\bibinfo {title} {Hierarchy of fractional {C}hern insulators and
  competing compressible states},\ }\href
  {https://doi.org/10.1103/PhysRevLett.111.126802} {\bibfield  {journal}
  {\bibinfo  {journal} {Phys. Rev. Lett.}\ }\textbf {\bibinfo {volume} {111}},\
  \bibinfo {pages} {126802} (\bibinfo {year} {2013})}\BibitemShut {NoStop}%
\bibitem [{\citenamefont {Abouelkomsan}\ \emph {et~al.}(2023)\citenamefont
  {Abouelkomsan}, \citenamefont {Yang},\ and\ \citenamefont
  {Bergholtz}}]{Abouelkomsan2023Quantum}%
  \BibitemOpen
  \bibfield  {author} {\bibinfo {author} {\bibfnamefont {A.}~\bibnamefont
  {Abouelkomsan}}, \bibinfo {author} {\bibfnamefont {K.}~\bibnamefont {Yang}},\
  and\ \bibinfo {author} {\bibfnamefont {E.~J.}\ \bibnamefont {Bergholtz}},\
  }\bibfield  {title} {\bibinfo {title} {Quantum metric induced phases in
  moir\'e materials},\ }\href
  {https://doi.org/10.1103/PhysRevResearch.5.L012015} {\bibfield  {journal}
  {\bibinfo  {journal} {Phys. Rev. Res.}\ }\textbf {\bibinfo {volume} {5}},\
  \bibinfo {pages} {L012015} (\bibinfo {year} {2023})}\BibitemShut {NoStop}%
\bibitem [{\citenamefont {Li}\ \emph {et~al.}(2010)\citenamefont {Li},
  \citenamefont {Luhman}, \citenamefont {Tsui}, \citenamefont {Pfeiffer},\ and\
  \citenamefont {West}}]{Li2010Observation}%
  \BibitemOpen
  \bibfield  {author} {\bibinfo {author} {\bibfnamefont {W.}~\bibnamefont
  {Li}}, \bibinfo {author} {\bibfnamefont {D.~R.}\ \bibnamefont {Luhman}},
  \bibinfo {author} {\bibfnamefont {D.~C.}\ \bibnamefont {Tsui}}, \bibinfo
  {author} {\bibfnamefont {L.~N.}\ \bibnamefont {Pfeiffer}},\ and\ \bibinfo
  {author} {\bibfnamefont {K.~W.}\ \bibnamefont {West}},\ }\bibfield  {title}
  {\bibinfo {title} {Observation of reentrant phases induced by short-range
  disorder in the lowest {L}andau level of
  ${\mathrm{al}}_{x}{\mathrm{ga}}_{1\ensuremath{-}x}\mathrm{As}/{\mathrm{al}}_{0.32}{\mathrm{ga}}_{0.68}\mathrm{As}$
  heterostructures},\ }\href {https://doi.org/10.1103/PhysRevLett.105.076803}
  {\bibfield  {journal} {\bibinfo  {journal} {Phys. Rev. Lett.}\ }\textbf
  {\bibinfo {volume} {105}},\ \bibinfo {pages} {076803} (\bibinfo {year}
  {2010})}\BibitemShut {NoStop}%
\bibitem [{\citenamefont {Liu}\ \emph {et~al.}(2012)\citenamefont {Liu},
  \citenamefont {Pappas}, \citenamefont {Shayegan}, \citenamefont {Pfeiffer},
  \citenamefont {West},\ and\ \citenamefont {Baldwin}}]{Liu2012Observation}%
  \BibitemOpen
  \bibfield  {author} {\bibinfo {author} {\bibfnamefont {Y.}~\bibnamefont
  {Liu}}, \bibinfo {author} {\bibfnamefont {C.~G.}\ \bibnamefont {Pappas}},
  \bibinfo {author} {\bibfnamefont {M.}~\bibnamefont {Shayegan}}, \bibinfo
  {author} {\bibfnamefont {L.~N.}\ \bibnamefont {Pfeiffer}}, \bibinfo {author}
  {\bibfnamefont {K.~W.}\ \bibnamefont {West}},\ and\ \bibinfo {author}
  {\bibfnamefont {K.~W.}\ \bibnamefont {Baldwin}},\ }\bibfield  {title}
  {\bibinfo {title} {Observation of reentrant integer quantum {H}all states in
  the lowest {L}andau level},\ }\href
  {https://doi.org/10.1103/PhysRevLett.109.036801} {\bibfield  {journal}
  {\bibinfo  {journal} {Phys. Rev. Lett.}\ }\textbf {\bibinfo {volume} {109}},\
  \bibinfo {pages} {036801} (\bibinfo {year} {2012})}\BibitemShut {NoStop}%
\bibitem [{\citenamefont {Te{\ifmmode \check{s}\else \v{s}\fi{}}anovi{\ifmmode
  \acute{c}\else \'{c}\fi{}}}\ \emph {et~al.}(1989)\citenamefont {Te{\ifmmode
  \check{s}\else \v{s}\fi{}}anovi{\ifmmode \acute{c}\else \'{c}\fi{}}},
  \citenamefont {Axel},\ and\ \citenamefont {Halperin}}]{Tesanovic1989Hall}%
  \BibitemOpen
  \bibfield  {author} {\bibinfo {author} {\bibfnamefont {Z.}~\bibnamefont
  {Te{\ifmmode \check{s}\else \v{s}\fi{}}anovi{\ifmmode \acute{c}\else
  \'{c}\fi{}}}}, \bibinfo {author} {\bibfnamefont {F.}~\bibnamefont {Axel}},\
  and\ \bibinfo {author} {\bibfnamefont {B.~I.}\ \bibnamefont {Halperin}},\
  }\bibfield  {title} {\bibinfo {title} {``{H}all crystal'' versus {W}igner
  crystal},\ }\href {https://doi.org/10.1103/PhysRevB.39.8525} {\bibfield
  {journal} {\bibinfo  {journal} {Phys. Rev. B}\ }\textbf {\bibinfo {volume}
  {39}},\ \bibinfo {pages} {8525} (\bibinfo {year} {1989})}\BibitemShut
  {NoStop}%
\bibitem [{\citenamefont {MacDonald}(1984)}]{MacDonald1984Influence}%
  \BibitemOpen
  \bibfield  {author} {\bibinfo {author} {\bibfnamefont {A.~H.}\ \bibnamefont
  {MacDonald}},\ }\bibfield  {title} {\bibinfo {title} {Influence of
  {L}andau-level mixing on the charge-density-wave state of a two-dimensional
  electron gas in a strong magnetic field},\ }\href
  {https://doi.org/10.1103/PhysRevB.30.4392} {\bibfield  {journal} {\bibinfo
  {journal} {Phys. Rev. B}\ }\textbf {\bibinfo {volume} {30}},\ \bibinfo
  {pages} {4392} (\bibinfo {year} {1984})}\BibitemShut {NoStop}%
\bibitem [{\citenamefont {Gon{\ifmmode \mbox{\c{c}}\else \c{c}\fi{}}alves}\
  \emph {et~al.}(2026)\citenamefont {Gon{\ifmmode \mbox{\c{c}}\else
  \c{c}\fi{}}alves}, \citenamefont {Mendez-Valderrama}, \citenamefont
  {Herzog-Arbeitman}, \citenamefont {Yu}, \citenamefont {Xu}, \citenamefont
  {Xiao}, \citenamefont {Bernevig},\ and\ \citenamefont
  {Regnault}}]{Goncalves2026Spinless}%
  \BibitemOpen
  \bibfield  {author} {\bibinfo {author} {\bibfnamefont {M.}~\bibnamefont
  {Gon{\ifmmode \mbox{\c{c}}\else \c{c}\fi{}}alves}}, \bibinfo {author}
  {\bibfnamefont {J.~F.}\ \bibnamefont {Mendez-Valderrama}}, \bibinfo {author}
  {\bibfnamefont {J.}~\bibnamefont {Herzog-Arbeitman}}, \bibinfo {author}
  {\bibfnamefont {J.}~\bibnamefont {Yu}}, \bibinfo {author} {\bibfnamefont
  {X.}~\bibnamefont {Xu}}, \bibinfo {author} {\bibfnamefont {D.}~\bibnamefont
  {Xiao}}, \bibinfo {author} {\bibfnamefont {B.~A.}\ \bibnamefont {Bernevig}},\
  and\ \bibinfo {author} {\bibfnamefont {N.}~\bibnamefont {Regnault}},\
  }\bibfield  {title} {\bibinfo {title} {Spinless and spinful charge
  excitations in moir\'e fractional {C}hern insulators},\ }\href
  {https://doi.org/10.1103/gmkp-bxf2} {\bibfield  {journal} {\bibinfo
  {journal} {Phys. Rev. Lett.}\ }\textbf {\bibinfo {volume} {136}},\ \bibinfo
  {pages} {196503} (\bibinfo {year} {2026})}\BibitemShut {NoStop}%
\bibitem [{\citenamefont {Yan}\ \emph {et~al.}(2026)\citenamefont {Yan},
  \citenamefont {Li}, \citenamefont {Soejima},\ and\ \citenamefont
  {Khalaf}}]{Yan2026Anyon}%
  \BibitemOpen
  \bibfield  {author} {\bibinfo {author} {\bibfnamefont {Z.}~\bibnamefont
  {Yan}}, \bibinfo {author} {\bibfnamefont {Q.}~\bibnamefont {Li}}, \bibinfo
  {author} {\bibfnamefont {T.}~\bibnamefont {Soejima}},\ and\ \bibinfo {author}
  {\bibfnamefont {E.}~\bibnamefont {Khalaf}},\ }\bibfield  {title} {\bibinfo
  {title} {Anyon dispersion in {A}haronov-{C}asher bands and implications for
  twisted ${\mathrm{mote}}_{2}$},\ }\href {https://doi.org/10.1103/shn5-vwgt}
  {\bibfield  {journal} {\bibinfo  {journal} {Phys. Rev. B}\ }\textbf {\bibinfo
  {volume} {113}},\ \bibinfo {pages} {235155} (\bibinfo {year}
  {2026})}\BibitemShut {NoStop}%
\bibitem [{\citenamefont {Schleith}\ \emph {et~al.}(2026)\citenamefont
  {Schleith}, \citenamefont {Soejima},\ and\ \citenamefont
  {Khalaf}}]{Schleith2026Anyon}%
  \BibitemOpen
  \bibfield  {author} {\bibinfo {author} {\bibfnamefont {M.-L.}\ \bibnamefont
  {Schleith}}, \bibinfo {author} {\bibfnamefont {T.}~\bibnamefont {Soejima}},\
  and\ \bibinfo {author} {\bibfnamefont {E.}~\bibnamefont {Khalaf}},\
  }\bibfield  {title} {\bibinfo {title} {Anyon dispersion from nonuniform
  magnetic field on the sphere},\ }\href {https://doi.org/10.1103/gzdq-wkxj}
  {\bibfield  {journal} {\bibinfo  {journal} {Phys. Rev. Lett.}\ }\textbf
  {\bibinfo {volume} {137}},\ \bibinfo {pages} {046506} (\bibinfo {year}
  {2026})}\BibitemShut {NoStop}%
\bibitem [{\citenamefont {Ouyang}\ and\ \citenamefont
  {Wu}(2026)}]{ouyang2026chiral}%
  \BibitemOpen
  \bibfield  {author} {\bibinfo {author} {\bibfnamefont {Y.}~\bibnamefont
  {Ouyang}}\ and\ \bibinfo {author} {\bibfnamefont {F.}~\bibnamefont {Wu}},\
  }\href {https://arxiv.org/abs/2610.40469} {\bibinfo {title} {Chiral $p$-wave
  superconductivity proximate to a fractional {C}hern insulator and a
  topological charge density wave in a moir\'e {C}hern band}} (\bibinfo {year}
  {2026}),\ \Eprint {https://arxiv.org/abs/2610.40469} {arXiv:2610.40469
  [cond-mat.supr-con]} \BibitemShut {NoStop}%
\bibitem [{\citenamefont {Xu}\ \emph {et~al.}(2025{\natexlab{b}})\citenamefont
  {Xu}, \citenamefont {Zou}, \citenamefont {Peshcherenko}, \citenamefont
  {Jahin}, \citenamefont {Li}, \citenamefont {Lin},\ and\ \citenamefont
  {Zhang}}]{Xu2025Chiral}%
  \BibitemOpen
  \bibfield  {author} {\bibinfo {author} {\bibfnamefont {C.}~\bibnamefont
  {Xu}}, \bibinfo {author} {\bibfnamefont {N.}~\bibnamefont {Zou}}, \bibinfo
  {author} {\bibfnamefont {N.}~\bibnamefont {Peshcherenko}}, \bibinfo {author}
  {\bibfnamefont {A.}~\bibnamefont {Jahin}}, \bibinfo {author} {\bibfnamefont
  {T.}~\bibnamefont {Li}}, \bibinfo {author} {\bibfnamefont {S.-Z.}\
  \bibnamefont {Lin}},\ and\ \bibinfo {author} {\bibfnamefont {Y.}~\bibnamefont
  {Zhang}},\ }\bibfield  {title} {\bibinfo {title} {Chiral superconductivity
  from spin polarized {C}hern band in twisted ${\mathrm{mote}}_{2}$},\ }\href
  {https://doi.org/10.1103/h22z-4hsj} {\bibfield  {journal} {\bibinfo
  {journal} {Phys. Rev. Lett.}\ }\textbf {\bibinfo {volume} {135}},\ \bibinfo
  {pages} {266005} (\bibinfo {year} {2025}{\natexlab{b}})}\BibitemShut
  {NoStop}%
\bibitem [{\citenamefont {Wang}\ and\ \citenamefont
  {Zaletel}(2025)}]{wang2025chiral}%
  \BibitemOpen
  \bibfield  {author} {\bibinfo {author} {\bibfnamefont {T.}~\bibnamefont
  {Wang}}\ and\ \bibinfo {author} {\bibfnamefont {M.~P.}\ \bibnamefont
  {Zaletel}},\ }\href {https://arxiv.org/abs/2507.07921} {\bibinfo {title}
  {Chiral superconductivity near a fractional {C}hern insulator}} (\bibinfo
  {year} {2025}),\ \Eprint {https://arxiv.org/abs/2507.07921} {arXiv:2507.07921
  [cond-mat.str-el]} \BibitemShut {NoStop}%
\bibitem [{\citenamefont {Levin}\ and\ \citenamefont
  {Stern}(2009)}]{Levin2009Fractional}%
  \BibitemOpen
  \bibfield  {author} {\bibinfo {author} {\bibfnamefont {M.}~\bibnamefont
  {Levin}}\ and\ \bibinfo {author} {\bibfnamefont {A.}~\bibnamefont {Stern}},\
  }\bibfield  {title} {\bibinfo {title} {Fractional topological insulators},\
  }\href {https://doi.org/10.1103/PhysRevLett.103.196803} {\bibfield  {journal}
  {\bibinfo  {journal} {Phys. Rev. Lett.}\ }\textbf {\bibinfo {volume} {103}},\
  \bibinfo {pages} {196803} (\bibinfo {year} {2009})}\BibitemShut {NoStop}%
\bibitem [{\citenamefont {Kwan}\ \emph {et~al.}(2026)\citenamefont {Kwan},
  \citenamefont {Wagner}, \citenamefont {Yu}, \citenamefont {Dagnino},
  \citenamefont {Jiang}, \citenamefont {Xu}, \citenamefont {Bernevig},
  \citenamefont {Neupert},\ and\ \citenamefont {Regnault}}]{Kwan2026Regarding}%
  \BibitemOpen
  \bibfield  {author} {\bibinfo {author} {\bibfnamefont {Y.~H.}\ \bibnamefont
  {Kwan}}, \bibinfo {author} {\bibfnamefont {G.}~\bibnamefont {Wagner}},
  \bibinfo {author} {\bibfnamefont {J.}~\bibnamefont {Yu}}, \bibinfo {author}
  {\bibfnamefont {A.~K.}\ \bibnamefont {Dagnino}}, \bibinfo {author}
  {\bibfnamefont {Y.}~\bibnamefont {Jiang}}, \bibinfo {author} {\bibfnamefont
  {X.}~\bibnamefont {Xu}}, \bibinfo {author} {\bibfnamefont {B.~A.}\
  \bibnamefont {Bernevig}}, \bibinfo {author} {\bibfnamefont {T.}~\bibnamefont
  {Neupert}},\ and\ \bibinfo {author} {\bibfnamefont {N.}~\bibnamefont
  {Regnault}},\ }\bibfield  {title} {\bibinfo {title} {Regarding the existence
  of abelian fractional topological insulators in twisted {M}o{T}e$_2$ and
  related systems},\ }\href {https://doi.org/10.1038/s42005-025-02483-6}
  {\bibfield  {journal} {\bibinfo  {journal} {Communications Physics}\ }\textbf
  {\bibinfo {volume} {9}},\ \bibinfo {pages} {52} (\bibinfo {year}
  {2026})}\BibitemShut {NoStop}%
\bibitem [{\citenamefont {Wang}\ \emph {et~al.}(2026)\citenamefont {Wang},
  \citenamefont {Minarik}, \citenamefont {Li}, \citenamefont {Kwan},
  \citenamefont {Yuan}, \citenamefont {Anderson}, \citenamefont {Hu},
  \citenamefont {Ingham}, \citenamefont {Choe}, \citenamefont {Taniguchi},
  \citenamefont {Watanabe}, \citenamefont {Roy}, \citenamefont {Chu},
  \citenamefont {Queiroz}, \citenamefont {Hone}, \citenamefont {Regnault},
  \citenamefont {Xu},\ and\ \citenamefont {Zhu}}]{Wang2026Candidate}%
  \BibitemOpen
  \bibfield  {author} {\bibinfo {author} {\bibfnamefont {Y.}~\bibnamefont
  {Wang}}, \bibinfo {author} {\bibfnamefont {G.~E.}\ \bibnamefont {Minarik}},
  \bibinfo {author} {\bibfnamefont {W.}~\bibnamefont {Li}}, \bibinfo {author}
  {\bibfnamefont {Y.}~\bibnamefont {Kwan}}, \bibinfo {author} {\bibfnamefont
  {S.}~\bibnamefont {Yuan}}, \bibinfo {author} {\bibfnamefont {E.}~\bibnamefont
  {Anderson}}, \bibinfo {author} {\bibfnamefont {C.}~\bibnamefont {Hu}},
  \bibinfo {author} {\bibfnamefont {J.}~\bibnamefont {Ingham}}, \bibinfo
  {author} {\bibfnamefont {J.}~\bibnamefont {Choe}}, \bibinfo {author}
  {\bibfnamefont {T.}~\bibnamefont {Taniguchi}}, \bibinfo {author}
  {\bibfnamefont {K.}~\bibnamefont {Watanabe}}, \bibinfo {author}
  {\bibfnamefont {X.}~\bibnamefont {Roy}}, \bibinfo {author} {\bibfnamefont
  {J.-H.}\ \bibnamefont {Chu}}, \bibinfo {author} {\bibfnamefont
  {R.}~\bibnamefont {Queiroz}}, \bibinfo {author} {\bibfnamefont {J.~C.}\
  \bibnamefont {Hone}}, \bibinfo {author} {\bibfnamefont {N.}~\bibnamefont
  {Regnault}}, \bibinfo {author} {\bibfnamefont {X.}~\bibnamefont {Xu}},\ and\
  \bibinfo {author} {\bibfnamefont {X.}~\bibnamefont {Zhu}},\ }\bibfield
  {title} {\bibinfo {title} {Candidate for a fractional topological insulator
  in twisted ${\mathrm{mote}}_{2}$},\ }\href
  {https://doi.org/10.1103/bvrb-z4hj} {\bibfield  {journal} {\bibinfo
  {journal} {Phys. Rev. X}\ }\textbf {\bibinfo {volume} {16}},\ \bibinfo
  {pages} {031009} (\bibinfo {year} {2026})}\BibitemShut {NoStop}%
\bibitem [{\citenamefont {Yu}\ \emph {et~al.}(2024)\citenamefont {Yu},
  \citenamefont {Herzog-Arbeitman}, \citenamefont {Wang}, \citenamefont
  {Vafek}, \citenamefont {Bernevig},\ and\ \citenamefont
  {Regnault}}]{Yu2024Fractional}%
  \BibitemOpen
  \bibfield  {author} {\bibinfo {author} {\bibfnamefont {J.}~\bibnamefont
  {Yu}}, \bibinfo {author} {\bibfnamefont {J.}~\bibnamefont
  {Herzog-Arbeitman}}, \bibinfo {author} {\bibfnamefont {M.}~\bibnamefont
  {Wang}}, \bibinfo {author} {\bibfnamefont {O.}~\bibnamefont {Vafek}},
  \bibinfo {author} {\bibfnamefont {B.~A.}\ \bibnamefont {Bernevig}},\ and\
  \bibinfo {author} {\bibfnamefont {N.}~\bibnamefont {Regnault}},\ }\bibfield
  {title} {\bibinfo {title} {Fractional {C}hern insulators versus nonmagnetic
  states in twisted bilayer {M}o{T}e$_2$},\ }\href
  {https://doi.org/10.1103/PhysRevB.109.045147} {\bibfield  {journal} {\bibinfo
   {journal} {Phys. Rev. B}\ }\textbf {\bibinfo {volume} {109}},\ \bibinfo
  {pages} {045147} (\bibinfo {year} {2024})}\BibitemShut {NoStop}%
\bibitem [{\citenamefont {Abouelkomsan}\ \emph {et~al.}(2024)\citenamefont
  {Abouelkomsan}, \citenamefont {Reddy}, \citenamefont {Fu},\ and\
  \citenamefont {Bergholtz}}]{Abouelkomsan2024Band}%
  \BibitemOpen
  \bibfield  {author} {\bibinfo {author} {\bibfnamefont {A.}~\bibnamefont
  {Abouelkomsan}}, \bibinfo {author} {\bibfnamefont {A.~P.}\ \bibnamefont
  {Reddy}}, \bibinfo {author} {\bibfnamefont {L.}~\bibnamefont {Fu}},\ and\
  \bibinfo {author} {\bibfnamefont {E.~J.}\ \bibnamefont {Bergholtz}},\
  }\bibfield  {title} {\bibinfo {title} {Band mixing in the quantum anomalous
  {H}all regime of twisted semiconductor bilayers},\ }\href
  {https://doi.org/10.1103/PhysRevB.109.L121107} {\bibfield  {journal}
  {\bibinfo  {journal} {Phys. Rev. B}\ }\textbf {\bibinfo {volume} {109}},\
  \bibinfo {pages} {L121107} (\bibinfo {year} {2024})}\BibitemShut {NoStop}%
\bibitem [{\citenamefont {Pescia}\ \emph {et~al.}(2022)\citenamefont {Pescia},
  \citenamefont {Han}, \citenamefont {Lovato}, \citenamefont {Lu},\ and\
  \citenamefont {Carleo}}]{PhysRevResearch.4.023138}%
  \BibitemOpen
  \bibfield  {author} {\bibinfo {author} {\bibfnamefont {G.}~\bibnamefont
  {Pescia}}, \bibinfo {author} {\bibfnamefont {J.}~\bibnamefont {Han}},
  \bibinfo {author} {\bibfnamefont {A.}~\bibnamefont {Lovato}}, \bibinfo
  {author} {\bibfnamefont {J.}~\bibnamefont {Lu}},\ and\ \bibinfo {author}
  {\bibfnamefont {G.}~\bibnamefont {Carleo}},\ }\bibfield  {title} {\bibinfo
  {title} {Neural-network quantum states for periodic systems in continuous
  space},\ }\href {https://doi.org/10.1103/PhysRevResearch.4.023138} {\bibfield
   {journal} {\bibinfo  {journal} {Phys. Rev. Res.}\ }\textbf {\bibinfo
  {volume} {4}},\ \bibinfo {pages} {023138} (\bibinfo {year}
  {2022})}\BibitemShut {NoStop}%
\bibitem [{\citenamefont {Li}\ \emph {et~al.}(2024{\natexlab{b}})\citenamefont
  {Li}, \citenamefont {Ye}, \citenamefont {Jiang}, \citenamefont {Wen},
  \citenamefont {Wang}, \citenamefont {Li}, \citenamefont {Li}, \citenamefont
  {He}, \citenamefont {Chen}, \citenamefont {Ren},\ and\ \citenamefont
  {Wang}}]{Li2024}%
  \BibitemOpen
  \bibfield  {author} {\bibinfo {author} {\bibfnamefont {R.}~\bibnamefont
  {Li}}, \bibinfo {author} {\bibfnamefont {H.}~\bibnamefont {Ye}}, \bibinfo
  {author} {\bibfnamefont {D.}~\bibnamefont {Jiang}}, \bibinfo {author}
  {\bibfnamefont {X.}~\bibnamefont {Wen}}, \bibinfo {author} {\bibfnamefont
  {C.}~\bibnamefont {Wang}}, \bibinfo {author} {\bibfnamefont {Z.}~\bibnamefont
  {Li}}, \bibinfo {author} {\bibfnamefont {X.}~\bibnamefont {Li}}, \bibinfo
  {author} {\bibfnamefont {D.}~\bibnamefont {He}}, \bibinfo {author}
  {\bibfnamefont {J.}~\bibnamefont {Chen}}, \bibinfo {author} {\bibfnamefont
  {W.}~\bibnamefont {Ren}},\ and\ \bibinfo {author} {\bibfnamefont
  {L.}~\bibnamefont {Wang}},\ }\bibfield  {title} {\bibinfo {title} {A
  computational framework for neural network-based variational {M}onte {C}arlo
  with {F}orward {L}aplacian},\ }\href
  {https://doi.org/10.1038/s42256-024-00794-x} {\bibfield  {journal} {\bibinfo
  {journal} {Nature Machine Intelligence}\ }\textbf {\bibinfo {volume} {6}},\
  \bibinfo {pages} {209} (\bibinfo {year} {2024}{\natexlab{b}})}\BibitemShut
  {NoStop}%
\bibitem [{\citenamefont {Sorella}(2005)}]{Sorella2005Wave}%
  \BibitemOpen
  \bibfield  {author} {\bibinfo {author} {\bibfnamefont {S.}~\bibnamefont
  {Sorella}},\ }\bibfield  {title} {\bibinfo {title} {Wave function
  optimization in the variational monte carlo method},\ }\href
  {https://doi.org/10.1103/PhysRevB.71.241103} {\bibfield  {journal} {\bibinfo
  {journal} {Phys. Rev. B}\ }\textbf {\bibinfo {volume} {71}},\ \bibinfo
  {pages} {241103(R)} (\bibinfo {year} {2005})}\BibitemShut {NoStop}%
\bibitem [{\citenamefont {Resta}(1998)}]{Resta1998Quantum}%
  \BibitemOpen
  \bibfield  {author} {\bibinfo {author} {\bibfnamefont {R.}~\bibnamefont
  {Resta}},\ }\bibfield  {title} {\bibinfo {title} {Quantum-mechanical position
  operator in extended systems},\ }\href
  {https://doi.org/10.1103/PhysRevLett.80.1800} {\bibfield  {journal} {\bibinfo
   {journal} {Phys. Rev. Lett.}\ }\textbf {\bibinfo {volume} {80}},\ \bibinfo
  {pages} {1800} (\bibinfo {year} {1998})}\BibitemShut {NoStop}%
\end{thebibliography}%

\clearpage

\appendix

\section{Variational ans\"atze and hyperparameters}
\label{Appendix:Ansatz}

Here we give the explicit form of the single-particle basis entering
Eq.~\eqref{eq:landau_orbitals} in the main text, the structure of the neural factors, and the
optimization protocol. Both ans\"atze share the same basis, the same neural
architecture and the same optimizer; they differ only in how the flux is
distributed.

\subsection{Landau level orbitals}

The determinant columns are expanded in the eigenstates of a \emph{uniform}
field carrying $N_\phi'$ flux quanta through the $L_x\times L_y$ torus, with
magnetic length $\ell'_B=\sqrt{L_xL_y/2\pi N_\phi'}$. In the Landau gauge,
$\bm A=(B'y,0)$ these are labeled by a Landau index $n$ and a guiding center
$k_m=2\pi m/L_x$ with $m=0,\dots,N_\phi'-1$,
\begin{equation}
\psi^{(N_\phi')}_{nk_m}(\bm r)=\frac{1}{\mathcal{N}_n}\sum_{s}
  e^{-ik_mx}\,e^{-isL_yx/\ell_B'^2}
  H_n\!\left(\frac{\tilde y_s}{\ell_B'}\right)e^{-\tilde y_s^2/2\ell_B'^2},
\label{eq:landau_basis}
\end{equation}
where $\tilde y_s=y-k_m\ell_B'^2-sL_y$, $H_n$ is a Hermite polynomial and
$\mathcal{N}_n=(2^nn!\,\ell_B' L_x\sqrt{\pi})^{1/2}$. The sum over images $s$
enforces periodicity in $y$ and the Gaussian makes it converge within a couple of
images. 

The coefficients $c^{n\bm k}$ of Eq.~\eqref{eq:landau_orbitals} in the mai text form a single
complex matrix $C$ of shape $(n_b N_\phi')\times N_e$, where
$n_b=n_{\text{max}}+1$ is the number of Landau levels retained. Every
determinant column is therefore a free complex combination of \emph{all}
$n_bN_\phi'$  basis functions. For the LSJ ansatz we reparameterize $C$
by a QR decomposition so that its columns stay orthonormal, which keeps the
determinant well conditioned as the orbitals localize.

\subsection{Composite-fermion factors}

\paragraph{Flux bookkeeping.} We are interested in $\nu=1/3$ and $\nu=2/3$. Since $\nu=n/(2pn \pm 1)$,  setting $p=1$ is sufficient. Writing $n$ for the
number of effective Landau levels the composite fermions fill, the determinant
carries $|N_\phi^\star| = N_e/n$ flux quanta and the flux-attachment factor
carries $N_\phi^{J} = 2pN_e$, with
\begin{equation}
N_\phi = N_\phi^{J} + N_\phi^\star ,
\end{equation}
equivalent to Eq.~\eqref{eq:CF_flux} in the main text. The two cases used here are
$\nu=1/3$: $n=1$, $|N_\phi^\star|=N_e$, $N_\phi = 2N_e+N_e = 3N_e$; and $\nu=2/3$: $n=2$, $|N_\phi^\star|=N_e/2$ with negative $N_\phi^\star$,
$N_\phi = 2N_e-N_e/2 = 3N_e/2$ .

\paragraph{Flux attachment.} The second line of Eq.~\eqref{eq:CF_Ansatz} in the main text is
evaluated as
\begin{equation}
\begin{split}
\Phi_{2p}(\{\bm r\}) = {}& e^{-\sum_i y_i^2/2\ell_J^2}
  \prod_{i<j}\vartheta_1(z_i-z_j|\tau)^{2p}\\
&\times\;\vartheta_{[\alpha,\beta]}\big(2pZ\,\big|\,2p\tau\big),
\end{split}
\label{eq:cf_jastrow}
\end{equation}
with $z_i=(x_i+iy_i)/L_x$, $Z=\sum\nolimits_i z_i$, $\tau=iL_y/L_x$ and
$\ell_J=\sqrt{L_xL_y/2\pi N_\phi^{J}}$ the magnetic length associated to the attached
flux.
The center-of-mass factor $\Theta_{\mathrm{CM}}$ is the theta function
$\vartheta_{[\alpha,\beta]}$ with characteristics $\alpha=r/2p$ and $\beta=0$;
it has $2p$ zeros in $Z$, and the integer $r=0,\dots,2p-1$ labels the $2p$
degenerate topological sectors. We take $r=0$ throughout.

\paragraph{Orbitals and conjugation.} The determinant columns use the same basis
\eqref{eq:landau_basis} with $N_\phi'=|N_\phi^\star|$ which corresponds to the enlarged magnetic
length $\ell_B^\star=\sqrt{L_xL_y/2\pi|N_\phi^\star|}$. At $\nu=2/3$ the
effective field seen by the composite fermions is reversed, so their orbitals
are the complex conjugates of Eq.~\eqref{eq:landau_basis}, anti-holomorphic
where $\Phi_{2p}$ is holomorphic. This is the $\phi^{\star}_a$ of
Eq.~\eqref{eq:CF_Ansatz}; since $\det[\bar\phi]=\overline{\det[\phi]}$ the
conjugation may equivalently be applied to the determinant as a whole.

\subsection{Neural factors}

Both the orbital prefactor $\eta_a$ and the Jastrow $J$ are read out from a
permutation-equivariant message-passing network of the type introduced in
Refs.~\cite{Pescia2024Message,Smith2024WC}. We use $3$ iterations of
attention-weighted message passing at width $32$ with GELU activations. The output is a one-body stream, $h_i$, for the vertices and a two-body stream, $h_{ij}$, for the edges. All inputs are
periodic \cite{PhysRevResearch.4.023138} --- $\sin$ and $\cos$ of the fractional pair separation together with
the minimum-image distance for the two-body stream, and $\sin$ and $\cos$ of the
fractional position for the one-body stream.

The orbital prefactor $\eta_a(\bm r_i;\{\bm r\})$ is a complex linear read-out of
$h_i$ and of the summed pair stream, $h_{ij}$. It multiplies the determinant columns and
therefore moves the nodes and introduces
correlations beyond a mean-field determinant. The Jastrow $\log J$ is read out from
$h_i$ alone and summed over electrons, which makes it symmetric and hence
nodeless; it is still all-body, since $h_i$ depends on every coordinate. In all runs reported here, the one-body stream retains absolute
positions, so the neural factors are not translation invariant and can support
symmetry broken states.

\subsection{Optimization and sampling}

Walkers are propagated with the Metropolis-adjusted Langevin algorithm (MALA), $5$
steps between samples, a step size retuned every $10$ iterations to hold the
acceptance ratio near its target and $100$ burn-in steps prior to training. All runs use
$1024$--$2048$ walkers and kinetic energies are evaluated with a forward-mode
Laplacian \cite{Li2024}. Parameters are optimized by stochastic reconfiguration \cite{Sorella2005Wave} with the SPRING algorithm. The learning rate decays as
\begin{equation}
    \eta(t)=\frac{\eta_0}{1+t/t_d},\qquad t_d=1000,
\end{equation}
and runs are optimized for $10^3$--$10^4$ steps.

The initial learning rate, $\eta_0$, is varied at every physical point along with the parameter initializations. We run several
values per point and report the lowest variational energy together with the
density of that run. Many of the points are transfer learned from a
converged neighbor in $B_1$ or $r_s$
rather than optimized from scratch for efficiency. Table~\ref{tab:fixed} collects the settings.


\begin{table}[htbp]
\caption{Settings held fixed across all production calculations. $n_b$ is the
number of Landau levels in Eq.~\eqref{eq:landau_orbitals} in the main text. The cap on the
update norm
is set equal to $\eta_0$, so the two are scanned together.}
\label{tab:fixed}
\renewcommand{\arraystretch}{1.2}
\begin{ruledtabular}
\begin{tabular}{lcc}
 & $\Psi_{\text{LSJ}}$ & $\Psi_{\text{CF}}$ \\
\hline
\multicolumn{3}{l}{\textit{Determinant}}\\
Landau levels $n_b$             & $1$--$6$      & $1$--$4$ \\
determinants                    & $1$     & $1$ \\
\hline
\multicolumn{3}{l}{\textit{Neural factors}}\\
message passing                 & \multicolumn{2}{c}{$3$ layers, width $32$, GELU} \\
\hline
\multicolumn{3}{l}{\textit{Sampling and optimization}}\\
sampler                         & \multicolumn{2}{c}{MALA, $5$ steps, adaptive $\tau$} \\
burn-in                  & \multicolumn{2}{c}{$100$} \\
walkers                         & \multicolumn{2}{c}{$1024$--$2048$} \\
optimizer                       & \multicolumn{2}{c}{SPRING} \\
proximal weight                 & \multicolumn{2}{c}{$0.9$} \\
step-size decay $t_d$           & \multicolumn{2}{c}{$1000$} \\
optimization steps              & \multicolumn{2}{c}{$10^3$--$10^4$} \\
\end{tabular}
\end{ruledtabular}
\end{table}


\subsection{Chern numbers}

Chern numbers are obtained from a flux-insertion pump. A twist $\theta_x$ is
threaded through the $x$ cycle of the torus and we follow the many-body
polarization along $y$ \cite{Resta1998Quantum},
\begin{equation}
Z_y(\theta_x)=\Big\langle\exp\Big(2\pi i\sum_j y_j/L_y\Big)\Big\rangle ,
\label{eq:resta}
\end{equation}
whose phase locates the charge centroid in units of $L_y$. The Chern number is
the winding of that phase over one flux quantum,
\begin{equation}
C=\frac{1}{2\pi}\sum_k\arg\big[Z_y(\theta_{k+1})/Z_y(\theta_k)\big],
\label{eq:winding}
\end{equation}
with each increment taken on the principal branch and $\theta_k$ a grid of $11$
equally spaced twists from $0$ to $2\pi$.

The pump must be adiabatic, so the twists are run as a chain: each is restarted
from the previous twist's optimized parameters \emph{and} its walkers, relaxed
for $600$ further steps. Two
diagnostics decide whether the winding may be read as a Chern number at all. The
energy must close around the loop, and does so to within $5\times10^{-5}$ Ha per
electron; and $|Z_y|$ must not drift, which it does by at most a few percent. A
state that fails either test has not returned to itself and its winding means
nothing. Since $\theta_x=0$ and $\theta_x=2\pi$ are optimized independently
rather than identified, Eq.~\eqref{eq:winding} is not quantized by construction,
and its departure from an integer is a further measure of convergence. If we do enforce these points are identical we get exactly $C=1$ for the hole crystal $C=0$ for all other crystals.

The electron and honeycomb crystals are pumped with no pinning potential; the
$\nu=2/3$ hole crystal slides under flux insertion and needs pinning ($0.3$ Ha
at the crystal sites) to stay commensurate. The measured windings are $0.02$--$0.076$ for the electron and honeycomb crystals, and $0.98$ for the hole
crystal, i.e. $C=0$ and $C=1$ respectively. Finally, the FCIs at $\nu=1/3$ and $\nu=2/3$ have Chern numbers $C=\nu$.
 
\section{Additional NQSVMC results}
\label{Appendix:AdditionalResults}
\begin{figure*}[htbp]
    \centering
    \includegraphics[width=0.95\textwidth]{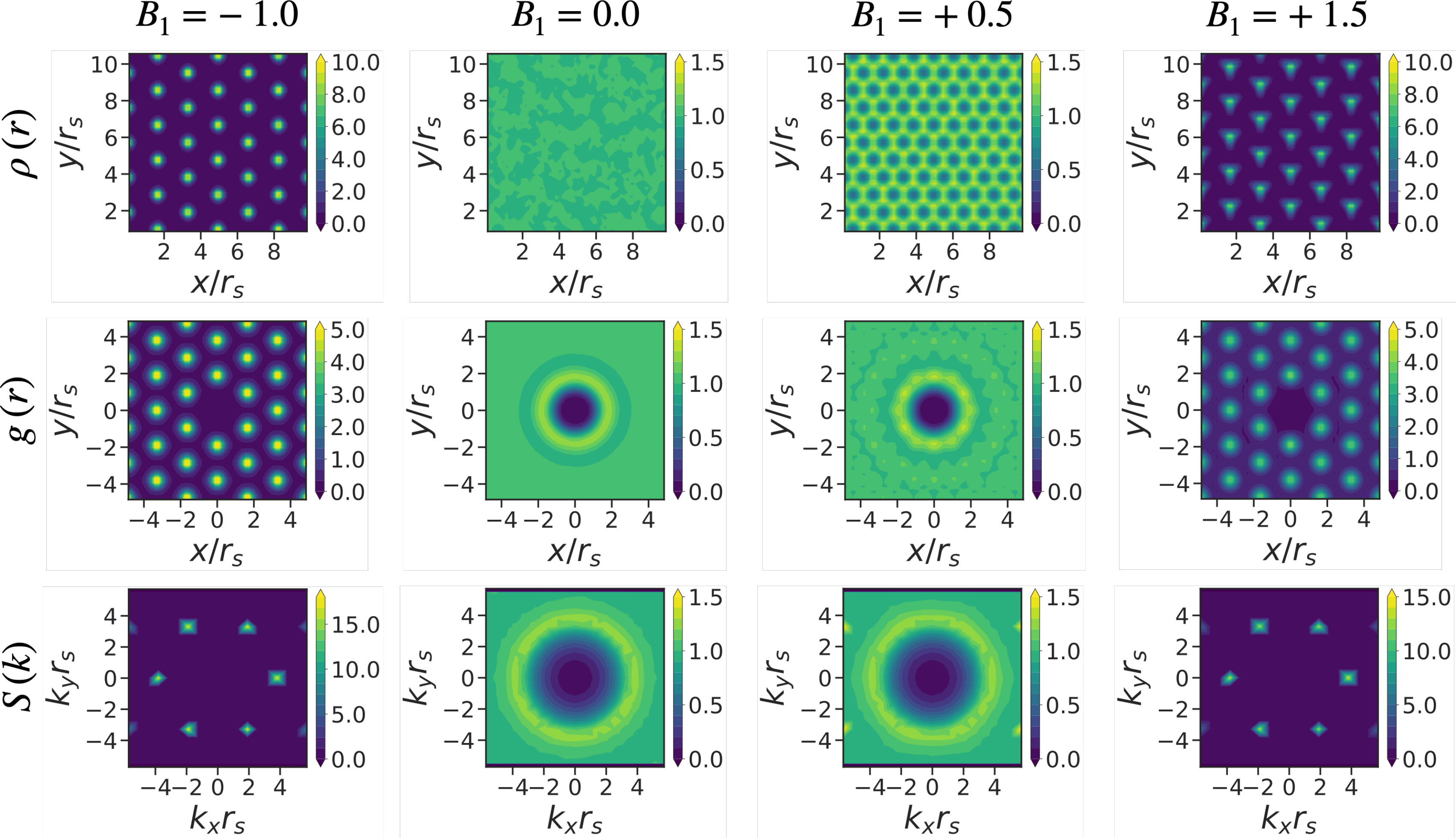}
    \caption{Observables for the ground states obtained at $\nu=1/3$ obtained from NQSVMC. The charge density $\rho({\bm r})$, correlation function $g({\bm r})$ and structure factor $S({\bm k})$ are shown for several values of $B_1$. These calculations correspond to $N_e=36$ electrons on a system containing $N=108$ magnetic field unit cells, and used $r_s=9.8$.}
    \label{fig:Observables_13}
\end{figure*}

In this section we present the additional observables that we used to characterize the ground state obtained from NQSVMC. Besides the charge densities, shown in the main text, we also compute the charge pair-correlation function $g({\bm r})$ and the charge structure factor $S({\bm k})$. The correlation function is defined as
\begin{align}
    g({\bm r})=\frac{\Omega}{N_e(N_e-1)}\int d^2 r^{\prime} \rho^{(2)}(\bm r',\bm r' + \bm r) ,
\end{align}
where $\Omega$ is the area of the simulation cell, and
\begin{equation}
    \rho^{(2)}(\bm r_1, \bm r_2) = \left \langle  \sum_{i\neq j } \delta (\bm r_1-r_i) \delta (\bm r_2-r_j) \right \rangle
\end{equation}
is the pair density. The structure factor is the Fourier transform of the correlation function, and is defined as

\begin{align}
    S(\bm k)=1+\frac{N_e-1 }{\Omega}\int d^2r\,e^{-i{\bm k}\cdot {\bm r}}\left[g({\bm r})-1 \right].
\end{align}
In Fig.~\ref{fig:Observables_13} we show observables for the different phases identified at filling $\nu=1/3$. All results were obtained for $r_s=9.8$. At $B_1=-1.0$ and $B_1=+1.5$, we obtain the $C_6$-crystal and $C_3$-crystal, respectively. The charge densities reveal the $\sqrt{3}\times\sqrt{3}$ charge order, with the localized orbitals having the corresponding symmetry. The correlation functions $g({\bm r})$ confirm crystalline order, which is supported by the Bragg peaks in $S({\bm k})$ emerging at the corners of the Brillouin zone defined by the periodicity of the magnetic field. We note that the magnitude in the peaks of $g({\bm r})$ and $S({\bm k})$ is larger for the $C_6$-crystal than for the $C_3$-crystal, which is in agreement with the former phase having more compact orbitals. At $B_1=0$ and $B_1=+0.5$ we obtain Laughlin-like liquids. For the Landau level Hamiltonian ($B_1=0$) the density is uniform, while for a generic AC Hamiltonian the density follows the magnetic field periodicity. $S({\bm k})$ and $g({\bm r})$ for both Laughlin-like states show liquid correlations. For $B_1\neq0$ we observe peaks at the reciprocal lattice vectors, reflecting the periodicity of the magnetic field.

\begin{figure*}[htbp]
    \centering
    \includegraphics[width=0.95\textwidth]{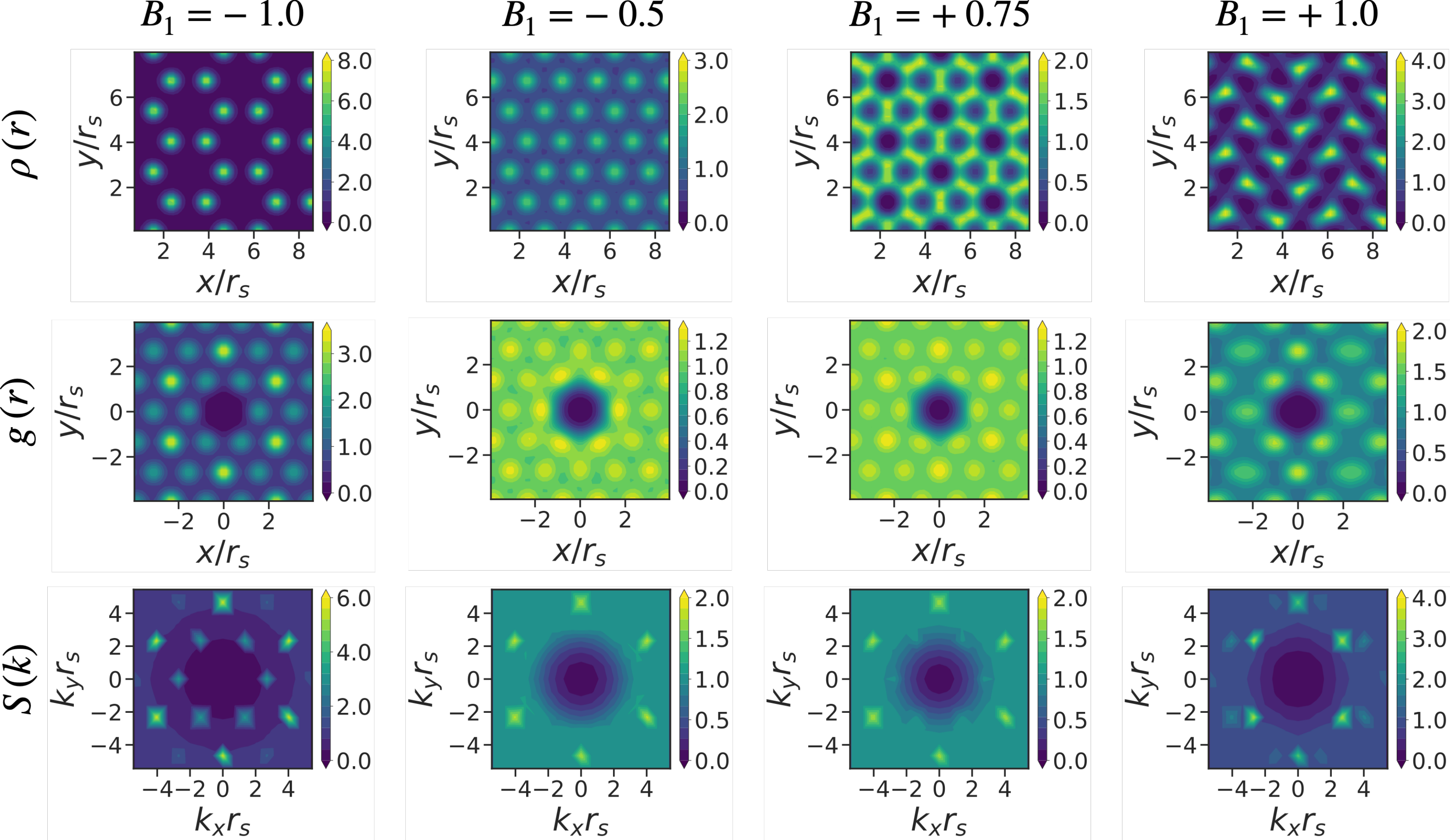}
    \caption{Observables for the ground states obtained at $\nu=2/3$ obtained from NQSVMC. The charge density $\rho({\bm r})$, correlation function $g({\bm r})$ and structure factor $S({\bm k})$ are shown for several values of $B_1$. These calculations correspond to $N_e=24$ electrons on a system containing $N=36$ magnetic field unit cells. The first, second, and fourth columns correspond to $r_s=6.92$ and the third column corresponds to $r_s=3.46$.}
    \label{fig:Observables_23}
\end{figure*}

In Fig.~\ref{fig:Observables_23} we show observables for the different phases identified at filling $\nu=2/3$. The two electron crystals and the FCI were obtained at $r_s=6.92$, and the hole crystal at $r_s=3.46$. The honeycomb electron GWC at $B_1=-1.0$ shows Bragg peaks at the corners of the Brillouin zone in the structure factor, consistent with a $\sqrt{3}\times\sqrt{3}$ charge order. For the incommensurate electron GWC at $B_1=+1.0$, both the correlations function and charge order show peaks that indicate crystallization. Due to incommensurability with the magnetic field and with the periodic boundary conditions of the simulation cell, we expect the observables of this phase to be dependent on the finite size geometry used for calculations.

Finally, the Jain-like FCI at $B_1=-0.5$ displays liquid-like correlations with structure factor peaks at the reciprocal lattice vectors, as expected. The observables for the triangular hole GWC at $B_1=+0.75$ are similar to those of the FCIs, since the electrons are on a liquid-like state. However, peaks in the structure factor at the Brillouin zone corners indicate spontaneous breaking of the discrete translation symmetry imposed by the magnetic field to form a $\sqrt{3}\times\sqrt{3}$ charge order.

\end{document}